\documentclass[%
 reprint,
superscriptaddress,
 amsmath,amssymb,
 aps,
]{revtex4-2}

\usepackage{graphicx}
\usepackage{dcolumn}
\usepackage{bm}
\usepackage{braket}
\usepackage[]{siunitx}
\newcommand*{\balancecolsandclearpage}{%
  \close@column@grid
  \cleardoublepage
  \twocolumngrid
}
\usepackage{comment}
\usepackage{multirow}

\newcommand{\beginsupplement}{%
        \setcounter{table}{0}
        \renewcommand{\thetable}{S\arabic{table}}%
        \setcounter{figure}{0}
        \renewcommand{\thefigure}{S\arabic{figure}}%
     }

\begin{document}

\preprint{APS/123-QED}

\title{Lattice dynamics of LiNb$_{\text{1-x}}$Ta$_{\text{x}}$O$_{\text{3}}$ solid solutions: Theory and experiment}

\author{Felix Bernhardt}
    \affiliation{These Authors contributed equally to this work}
    \affiliation{Institut für Theoretische Physik, Center for Materials Research (LaMa), Justus-Liebig-Universität Gießen, 35392 Gießen, Germany}
\author{Soham Gharat}
    \affiliation{These Authors contributed equally to this work}
    \affiliation{Institut für Angewandte Physik, Technische Universität Dresden, 01062 Dresden, Germany}
\author{Alexander Kapp}
    \affiliation{These Authors contributed equally to this work}
    \affiliation{Institut für Theoretische Physik, Center for Materials Research (LaMa), Justus-Liebig-Universität Gießen, 35392 Gießen, Germany}
\author{Florian Pfeiffer}
    \affiliation{Institut für Theoretische Physik, Center for Materials Research (LaMa), Justus-Liebig-Universität Gießen, 35392 Gießen, Germany}
\author{Robin Buschbeck}
    \affiliation{Institut für Angewandte Physik, Technische Universität Dresden, 01062 Dresden, Germany}
\author{Franz Hempel}
    \affiliation{Institut für Angewandte Physik, Technische Universität Dresden, 01062 Dresden, Germany}
\author{Oleksiy Pashkin}
    \affiliation{Helmholtz-Zentrum Dresden-Rossendorf, 01328 Dresden, Germany}
\author{Susanne C. Kehr}
    \affiliation{Institut für Angewandte Physik, Technische Universität Dresden, 01062 Dresden, Germany}
\author{Michael Rüsing}
    \affiliation{Institut für Angewandte Physik, Technische Universität Dresden, 01062 Dresden, Germany}
    \affiliation{Paderborn University, Integrated Quantum Optics, Institute for Photonic Quantum Systems (PhoQS), Warburger Str. 100, 33098 Paderborn, Germany}
\author{Simone Sanna}
    \affiliation{Institut für Theoretische Physik, Center for Materials Research (LaMa), Justus-Liebig-Universität Gießen, 35392 Gießen, Germany}
\author{Lukas M. Eng}
    \affiliation{Institut für Angewandte Physik, Technische Universität Dresden, 01062 Dresden, Germany}
    \affiliation{ct.qmat: Dresden-Würzburg Cluster of Excellence—EXC 2147, TU Dresden, 01062 Dresden, Germany}

\date{\today}

\begin{abstract}
Lithium niobate (LNO) and lithium tantalate (LTO) see widespread use in fundamental research and commercial technologies reaching from electronics over classical optics to integrated quantum communication. In recent years, the mixed crystal system lithium niobate tantalate (LNT) allows for the dedicate engineering of material properties by combining the advantages of the two parental materials LNO and LTO. Vibrational spectroscopies such as Raman spectroscopy or (Fourier transform) infrared spectroscopy are vital techniques to provide detailed insight into the material properties, which is central to the analysis and optimization of devices. In this work, we present a joint experimental-theoretical approach allowing to unambiguously assign the spectral features in the LNT material family through both Raman and IR spectroscopy, as well as to provide an in-depth explanation for the observed scattering efficiencies based on first-principles calculations. The phononic contribution to the static dielectric tensor is calculated from the experimental and theoretical data using the generalized Lyddane-Sachs-Teller relation and compared with the results of the first-principles calculations. The joint methodology can be readily expanded to other materials and serves, e.g., as the basis for studying the role of point defects or doping.
\end{abstract}

\maketitle

\section{Introduction}
Lithium niobate (LiNbO$_3$, LNO) and lithium tantalate (LiTaO$_3$, LTO) are two isomorphous ferroelectrics (space group $R3c$) and are among the most widely used materials in electro-optic applications \cite{Weis85,Raeuber78,Volk08}. Both materials feature a ferroelectric to paraelectric structural phase transition, which, according to the actual knowledge, is a phase transition of second order \cite{6306010,Phillpot04}. Correspondingly, the spontaneous polarization $P_{S}$ steadily increases with decreasing temperature from 0 at the Curie temperature to a value of 71(62)\,$\mu$C/cm$^2$ and 60(55)\,$\mu$C/cm$^2$ for congruent (nearly stoichiometric) LNO \cite{Chen01} and LTO \cite{Kitamura98} at low temperatures, respectively. LNO is characterized by unusually large pyroelectric, piezoelectric, electro-optic, and photo-elastic coefficients \cite{Weis85}. The magnitude of these coefficients is less pronounced in LTO, which features, however, higher thermal stability due to, for example, lower thermally activated conductivity, which is favorable for high temperature sensing applications \cite{Yakhnevych2023}. In contrast, however, LTO has a much lower Curie temperature compared to LNO 874-958\,K for LTO \cite{Chen01, Nakamura08} vs. 1413-1475\,K for LNO \cite{Kitamura98, Kim03}), limiting its applications to lower temperature regions. Lithium niobate and lithium tantalate are birefringent, have useful acoustic wave properties \cite{TCLee03} and a rather large acousto-optic figure-of-merit. The wealth of physical effects and, more important, their magnitude, render LNO and LTO ideal candidates for acoustic and optical applications, exploiting both their bulk and surface properties \cite{Sanna17surf}. Recently, lithium-niobate-tantalate solid solutions (LiNb$_{1-x}$Ta$_x$O$_3$, LNT) have gained attention, as they conjugate the favorable properties of LNO with the thermal stability of LTO \cite{Yakhnevych2023,Bashir2023,ROSHCHUPKIN2023,Suhak2021}. Futhermore, LNT solid solutions allow for the tailoring of many material properties by adjusting the niobium-tantalum ratio and/or the  Li concentration \cite{Suhak2021,Bartasyte2019,ROSHCHUPKIN2023,Bashir2023,Wood08}. 

For a complete characterization regarding the niobium/tantalum ratio of the LNT solid solution, the knowledge of the dynamical properties of the crystal lattice of the end compounds LNO and LTO is crucial. Indeed, it bears information concerning, e.g., crystal symmetry, phase transition dynamics, and many other aspects. In this context, Raman spectroscopy, as well as the complementary infrared spectroscopy (IR) both provide access to study the dynamical properties of the lattice and, in turn, to characterize many fundamental material parameters \cite{RuesPRB,Bartasyte2019}.


As an example, the presence of one or more optical phonon modes that become soft close to the Curie temperature, is a typical signature of displacive type transitions. No soft modes exist in order-disorder type transitions, instead. Therefore, many different studies did focus on the investigation of the phonon modes of LNO and LTO \cite{SimoRaman15,Friedrich16,Caciuc01,Rapitis88,Samuel12,Friedrich15,Ridah97}. These investigations lead to the assignment of the experimentally detected spectral signatures associated to atomic displacement patterns. However, the data from the literature present a not fully consistent picture. In particular, in the case of LNO, there have been many ambiguities in the assignment of spectral features \cite{Hermet2007,Margueron2012,RuesPRB,SimoRaman15}. Some of the investigations, including Rayleigh scattering, Raman spectroscopy, and infrared reflectivity, demonstrated the existence of $A_1$-TO optical phonons becoming soft at high temperatures, suggesting a displacive nature of the transition \cite{Barker67,Johnston68,Servoin79}. Yet, no mode softening could be observed in other studies, including neutron and Raman scattering experiments, suggesting the order-disorder nature of the ferroelectric to paraelectric phase transition \cite{Chowdhury74,Kojima99,Penna76}. 
Moreover, the knowledge of the dynamical properties of the solid solutions is 
rather limited, despite decades of research \cite{SimoMix,RuesPRB,Bartasyte2019}.

Here, we investigate the dynamical properties of LiNb$_{1-x}$Ta$_x$O$_3$ 
solid solutions and their end compounds both theoretically and experimentally. Longitudinal-optical (LO) and transverse-optical (TO) phonon frequencies, IR spectra - more precisely the complex dielectric function - and Raman
spectra are calculated from first-principles within density functional theory and compared to experimental data obtained by both Fourier-transform infrared spectroscopy (FTIR) and $\mu$-Raman spectroscopy. Raman and FTIR spectroscopy are complementary techniques that, relying on different physical processes, give access to phonon modes of different symmetry in some materials, or in the case of LNO and LTO enable a much improved assignment of phonons, as some have a very low scattering efficiency in either Raman or IR. Calculated and experimentally observed phonon modes are compared to each other and, where existing, with available 
literature data. This allows for the assignment and determination of all optical phonon modes, including some that are difficult to measure solely by Raman or (FT)IR spectroscopies. The similarities and differences of the dynamical properties of LNO, LTO, and LNT are briefly discussed. Finally, the measured phonon frequencies are employed to estimate the phononic contribution to the dielectric tensor with the Lyddane-Sachs-Teller (LST)
relation \cite{LST1,CHAVES1973865}.

\section{Experiment}
\subsection{Samples}
In this work, we have investigated three types of single crystals: 1) LNO, 2)  LTO, and 3)  LNT mixed crystals with a tantalum concentration of $x = 0.70 \pm 0.03$. All crystals are grown at the congruent point, i.e. all crystals feature the typical lithium deficiency of approximately 1.5~mol\%, which is a result of the crystal growth technique \cite{Volk08}. The crystals are otherwise undoped. The LNO and LTO crystals are commercially obtained (Impex HighTech GmbH, Münster, Germany), while the LNT crystals were grown by the Czochralski method at the Leibniz Institute for Crystal Growth, IKZ, Berlin. The growth of the LNT crystals is described elsewhere \cite{Bashir2023}. The crystal orientation has been determined via X-ray diffraction and samples hasve been prepared as x-cut, y-cut, or z-cut, which allows to study all optically active phonon branches, as well as transverse- and longitudinal-polarized phonon modes. All crystal surfaces, which have been studied in Raman or FTIR spectroscopy, have been polished to optical quality.

\subsection{Raman spectroscopy}

Raman spectroscopy was carried out on a  LabRAM HR Evolution Raman spectroscope by Horiba Seisakusho. Excitation light is provided by a linearly polarized 17\,mW HeNe-laser at 632.8\,nm. The laser light is focused via an objective lens with a low numerical aperture of 0.3. The low numerical aperture allows one to average the signal over a larger sample volume, and it reduces the influence of focusing-related depolarization \cite{Saito2008,Spychala2020}. The scattered light is collected in back-reflection geometry via the same objective lens. The elastically scattered light is blocked via an appropriate edge filter and spectral analysis is performed in an attached spectrometer with a 600\,grooves/mm grating and detected with a charge-coupled device. The setup reaches a spectral resolution better than 2\,cm$^{-1}$. The light polarization in excitation and detection paths is set by automatized half-wave plates and linear polarizers, respectively. Calibration was done by measuring (100)-cut single crystalline silicon, which shows one prominent and sharp peak at 521\,cm$^{-1}$. More details on the setup can be found elsewhere \cite{Reitzig2021,Reitzig2022}.

For each sample and available crystal orientation, polarized Raman spectra are recorded by placing the focus more than 10~$\mu$m below the surface to limit spectroscopic signatures from the surface \cite{Sanna2011}. Due to the high Raman scattering efficiency in the LNT crystal family, high signal-to-noise ratio Raman spectra, which require no further smoothing, can be obtained with integration times of a few seconds at most. The obtained spectra are normalized to their respective maximum after the dark counts have been subtracted. No further data processing was performed for joint plotting with the calculated spectra. To accurately extract the peak frequencies, the spectra were fitted with Lorentzian functions.

\subsection{FTIR spectroscopy}

Reflectance FTIR spectroscopy is performed on a Bruker VERTEX 80v FTIR spectrometer in the range from 50 to about 1000\,cm$^{-1}$, covering the known phononic frequencies of LNO and LTO \cite{SimoRaman15}. The light is focused and collected via two ellipsoidal mirrors, which are aligned to a common focus. The spectrometer yields a spectral resolution of about 4\,cm$^{-1}$. Reference reflectance spectra are taken on a gold reference mirror. The light polarization of the glowbar source is set linear with a Tydex thin-film wire-grid polarizer.

\begin{figure}
    \centering
    \includegraphics[scale=0.5]{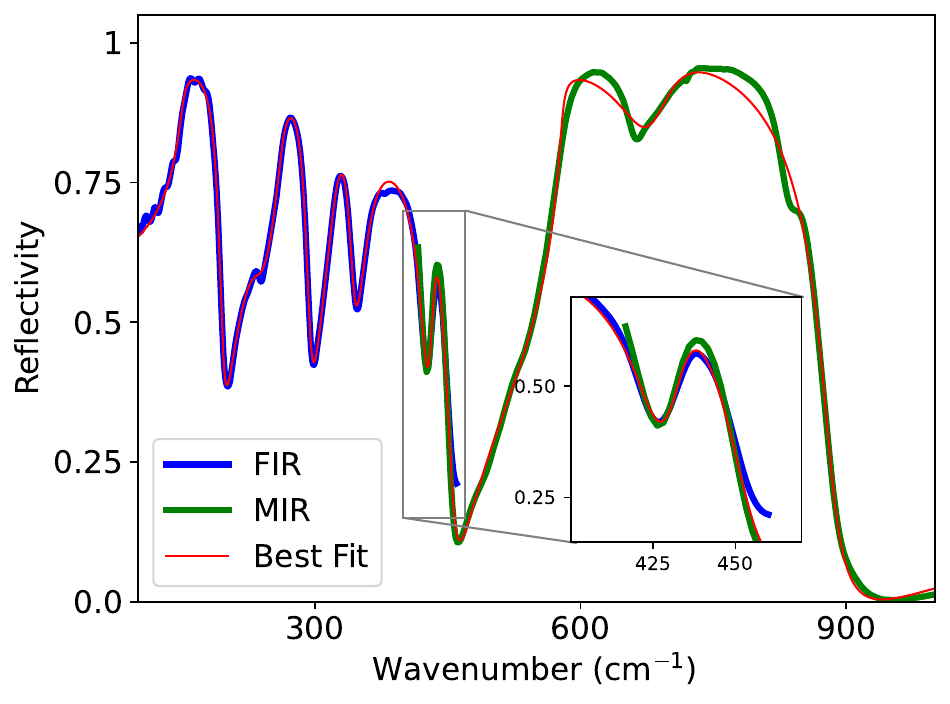}
    \caption{Exemplary FTIR-reflectance spectrum of the z-cut LNO sample in ordinary light polarization ($E \parallel x$). Due to limitations in the experimental setup, the spectra are measured with two different beam splitter/detector combinations, one operating in the far-infrared (FIR) region displayed in blue and one operating in the mid-infrared range (MIR) displayed in green. The spectra show a good overlap in the 400-450\,cm$^{-1}$ range, which is used to merge the two datasets. To extract the full dielectric function in the phononic range, the spectrum is fitted with an oscillator model as shown in Eq.~\ref{eq:four_parameter}. The best fit for the given spectrum is shown in red.}
    \label{fig:FTIR_raw}
\end{figure}

Figure \ref{fig:FTIR_raw} shows an exemplary raw data set taken on z-cut LNO, where the light is polarized along the extraordinary axis of the crystal ($E \parallel x$). Due to the limited spectral range of the available beam splitters, two different experimental settings for the mid-infrared (MIR; 400-1000\,cm$^{-1}$, Ge/KBr beam splitter, MIR-BLATGS detector), and far-infrared (FIR, 50-600\,cm$^{-1}$, Myloo beam splitter, FIR-DTGS detector) were used to measure the full range of phonon frequencies in the LNT crystal family. This can be readily seen in the raw data in Fig.~\ref{fig:FTIR_raw}, where the FIR data is displayed in blue and the MIR data in green. The two ranges show a good overlap in the 400-450\,cm$^{-1}$ range, as displayed in the inset, which is used to merge the datasets.

To obtain the dielectric function $\varepsilon(\omega)$ in the phononic frequency range, the obtained reflectance spectrum is fitted with a four parameter oscillator model of the form

\begin{equation}
\label{eq:four_parameter}
\varepsilon(\omega)=
\varepsilon_\infty\prod_{j=1}^{n}
\frac{\omega^2_{\mathrm{LO}j}-\omega^2+i\omega\gamma_{\mathrm{LO}j}}{\omega^2_{\mathrm{TO}j}-\omega^2+i\omega\gamma_{\mathrm{TO}j}}.
\end{equation}

In this equation the $4n+1$ independent (fitting) parameters are  $\omega_{\mathrm{LO}j}$ and $\omega_{\mathrm{TO}j}$, which refer to the frequencies of the $j$-th longitudinal optical (LO) and transverse-optical (TO) phonon modes, respectively, and $\gamma_{\mathrm{LO}j}$ and $\gamma_{\mathrm{TO}j}$, which refer to their respective damping constants, as well as the high frequency permittivity $\varepsilon_\infty$, which describes the dielectric contribution of the electronic background, which becomes dominant at higher frequencies (towards optical frequencies).

Compared to a more traditional three-parameter \emph{Drude-Lorentz} oscillator model, which features only one damping constant and was used in the past for LNO or LTO \cite{Barker1970}, fitting the four parameter model was chosen, because it describes the dielectric functions of materials with a strong LO-TO splitting commonly seen in polar crystals, like ferroelectrics, more accurately \cite{Gervais_1974a,Gervais1974b}. Furthermore, in the limit of $\omega \rightarrow 0$ equation \eqref{eq:four_parameter} is identical to the Lyddane-Sachs-Teller relation (LST) relation in Eq.~\eqref{eq:LST}.

The fitting procedure is performed via the software RefFIT \cite{Kuzmenko2005}, yielding the real $\varepsilon'$ and imaginary $\varepsilon''$ contributions of the dielectric function. A fitting result is displayed in Fig.~\ref{fig:FTIR_raw} which shows good agreement with the raw data. Subsequently in this work, only the extracted real $\varepsilon'$ and imaginary $\varepsilon''$ dielectric function are shown. Prior to fitting of each datasets, the number of expected phonons is set. Approximate starting frequencies are based on literature or Raman data. Specifically, in datasets containing A$_1$-symmetry phonons, we did set four phonons of A$_1$ type. However, for E-type symmetry phonons we only used eight instead of nine independent phonons, because from previous work it is known that the E-TO$_6$ has a very weak IR oscillator strength and is spectrally very close (10~cm$^{-1}$) to the more intense E-TO$_5$ \cite{Margueron2012,RuesPRB,SimoMix}. This is reversed in Raman spectroscopy, where the E-TO$_5$ is quasi-Raman silent and dominated by the E-TO$_6$, as also confirmed by theoretical calculations (Sec. VB).

\subsection{Phonon modes and Selection rules}

LNO and LTO are isostructural with two formula units per unit cell. This results in 27 optical phonon modes. Based on group theory, the optical phonon modes at the Brillouin zone center $\Gamma$  are further subdivided in four modes of A$_1$-type, five modes of A$_2$-type and nine doubly degenerate modes of E-type symmetry. Only the A$_1$ and E-symmetry phonons are Raman \emph{and} IR-active, while the A$_2$ phonons are optically silent. Furthermore, each phonon may be associated with a transverse optical (TO) frequency and a longitudional optical (LO) frequency, as discussed for Eq.~\eqref{eq:four_parameter}, which define the Reststrahlenband in the reflectance spectra. Therefore, to characterize the full optical response either by Raman or IR spectroscopy, four A$_1$-TO and four A$_1$-LO frequencies, as well as nine E-TO and nine E-LO frequencies need to be known. In the LNO crystal system it is common to label the phonons in ascending order, e.g. E-LO$_6$ refers to the sixth highest frequency E-type symmetry LO phonon.

As discussed below (Sec. III A), the mixed crystals cannot be described by the same unit cell due to the larger supercell necessary to take into account the arbitrary Nb-Ta content of the mixed crystals. Assuming the natural randomness in the Nb-Ta distribution, a large number of phonons will be present at the Brillouin-zone center leading to a natural broadening of peaks and a lifting of the selection rules. Therefore, strictly speaking the selection rules true for the end compounds LNO and LTO are not necessarily valid for the mixed crystals. Nevertheless, we chose to label the features in the mixed crystals only to a single mode in accordance with its closest phononic analogue in the end compound. These quasi A$_1$ and E-modes are labeled with a prime, e.g. A$'_1$-TO$_1$ denoting the peak that is similar to the A$_1$-TO$_1$ mode in either the LNO or LTO end compounds.

The selection rules in Raman spectroscopy can be calculated based on the known Raman tensors, which summarizes, both, the propagation direction of the phonons to distinguish LO and TO phonons, and assign which phonons for a given set of excitation and detection light polarization can be detected. The tensors can be found elsewhere \cite{RuesPRB,Bartasyte2012,Margueron2012,Bartasyte2019}. The scattering geometry in Raman spectroscopy is described with Porto's notation: $k_i(e_i,e_s) k_s$. The vectors $k_i$ and $k_s$ mark the direction of the incident and scattered light in crystal coordinates, respectively, while $e_i$ and $e_s$ label its electric field orientation, i.e. the polarization, also in crystal coordinates. Table \ref{tab:porto} summarizes the twelve possible scattering geometries in back-scattering Raman spectroscopy, i.e. $k_i = -k_s = \overline{k_i}$.

\begin{table}
	\begin{ruledtabular}
		\caption{\label{tab:porto} Observable phonon modes and Raman tensor elements recorded in back-scattering configuration for LNO and LTO single crystals. The tensor elements are defined as in Ref. \cite{RuesPRB}.}
		\begin{tabular}{cccc}
			Scattering & Symmetry  &  \multicolumn{2}{c}{Tensor element}\\
			configuration & species  & TO & LO\\
			\hline
			x(y,y)$\overline{\rm{x}}$ & $A_1$-TO, $E$-TO  & $a^2+c^2$   & \\
			x(y,z)$\overline{\rm{x}}$ & $E$-TO            & $d^2$ & \\
			x(z,y)$\overline{\rm{x}}$ & $E$-TO            & $d^2$ & \\
			x(z,z)$\overline{\rm{x}}$ & $A_1$-TO          & $b^2$ & \\
   \hline
			y(x,x)$\overline{\rm{y}}$ & $A_1$-TO, $E$-LO  & $a^2$ & $c^2$ \\
			y(x,z)$\overline{\rm{y}}$ & $E$-TO            & $d^2$ & \\
			y(z,x)$\overline{\rm{y}}$ & $E$-TO            & $d^2$ & \\
			y(z,z)$\overline{\rm{y}}$ & $A_1$-TO          & $b^2$ & \\
   \hline
			z(x,x)$\overline{\rm{z}}$ & $A_1$-LO, $E$-TO  & $c^2$ & $a^2$ \\
			z(x,y)$\overline{\rm{z}}$ & $E$-TO            & $c^2$ & \\
			z(y,x)$\overline{\rm{z}}$ & $E$-TO            & $c^2$ & \\
			z(y,y)$\overline{\rm{z}}$ & $A_1$-LO, $E$-TO  & $c^2$ & $a^2$\\
		\end{tabular}
	\end{ruledtabular}
\end{table}

The selection rules in (FT)IR spectroscopy are more straightforward. Here, A$_1$-type symmetry phonons are excited, when the incident and scattered light is polarized parallel to the extraordinary axis (z-axis) of the crystal, i.e. $E \parallel z$. In contrast, the E-type phonon contributions to the dielectric response are observed when the incident light is polarized parallel to the ordinary crystal axis (xy-plane), i.e. $E \parallel x$ or $E \parallel y$. These selection rules are summarized in Tab. \ref{tab:FTIR}. Note, that the dielectric tensor $\varepsilon$ of the LNT crystal family has only two independent components.

\begin{table}
	\begin{ruledtabular}
		\caption{\label{tab:FTIR} Observable phonon modes recorded for light polarization (electric field) with respect to the crystal axes of LNO or LTO.}
		\begin{tabular}{cc}
			Light polarization (E-field) & Symmetry species \\ \hline
			$E \parallel z$ & $A_1$ \\ 
			$E \parallel y$ & $E$  \\
			$E \parallel x$ & $E$ 
		\end{tabular}
	\end{ruledtabular}
\end{table}

\section{Density functional perturbation theory}
\subsection{Computational details}
Phonon eigenmodes, eigenfrequencies and effective charges are calculated within density functional perturbation theory (DFPT) as implemented in VASP \cite{Kresse1993,Kresse1996,Kresse1996_2}, and post processed by Phonopy \cite{Togo2023,Togo2023_2}, an open source python package created for performing basic phonon calculations at the harmonic level.
The dielectric function is calculated within the independent particle approximation (IPA) using the same software. Projector augmented wave (PAW) potentials \cite{Bloechl94,Joubert1999} with exchange-correlation functional in the formulation of Perdew-Burke Ernzerhof revised for solids (PBEsol) \cite{Perdew1996,Perdew2008} and electronic configurations 1s$^2$2s$^1$, 4s$^2$4p$^6$4d$^3$5s$^2$, 
5p$^6$5d$^3$6s$^2$, and 2s$^2$2p$^4$ for Li, Nb, Ta, and O, respectively, are used. 
Rhombohedral, primitive unit cells with $R3c$ symmetry model the ferroelectric phase of LNO and LTO, respectively. The used equilibrium lattice parameters are $a_{\mathrm{LT}}$=5.474\,{\AA} and $\alpha_{\mathrm{LT}}$=56.171$^\circ$ for LTO, as well as $a_{\mathrm{LN}}$=5.494\,{\AA} and $\alpha_{\mathrm{LN}}$=55.867$^\circ$ for LNO. These values are in close agreement with the measured lattice parameters \cite{RuesPRB,Bartasyte2012,Huband2017}. The ionic positions are optimized, such that all forces acting on the ions are lower than 0.005\,eV/{\AA}. A $8\times8\times8$ Monkhorst-Pack K-point mesh \cite{Pack1977}, as well as an energy cutoff at 500\,eV are needed to converge electronic energies to 10\,meV. A Gaussian smearing with width 0.05\,eV is applied to the Fermi occupancies. 46 conduction bands are considered to obtain a dielectric function converged up to 10\,eV.

\begin{figure}
    \includegraphics[scale=0.5]{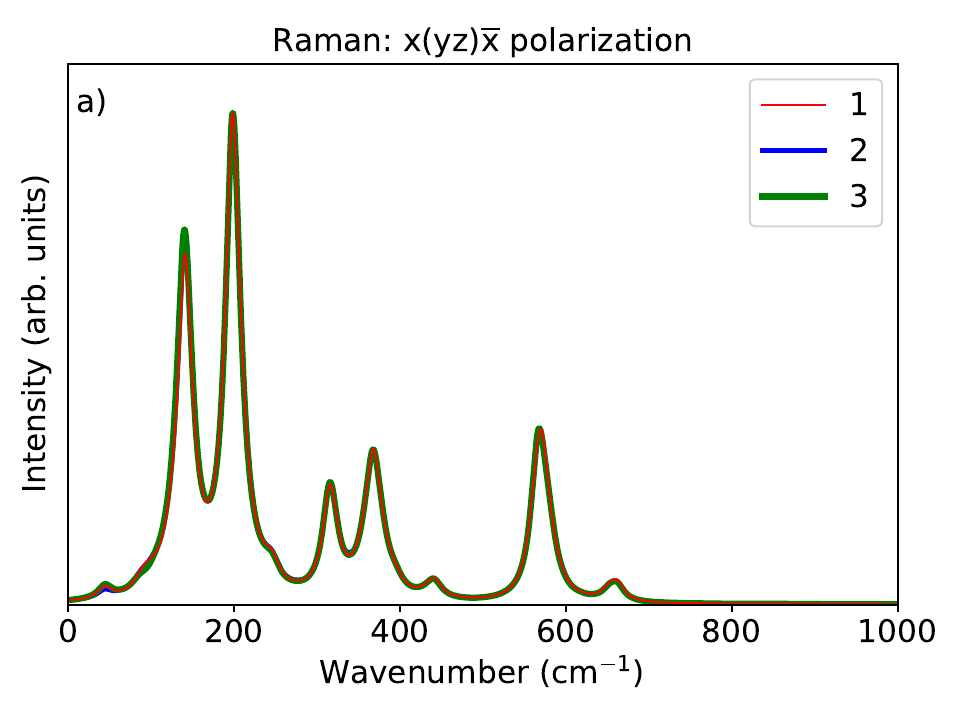}
    \includegraphics[scale=0.5]{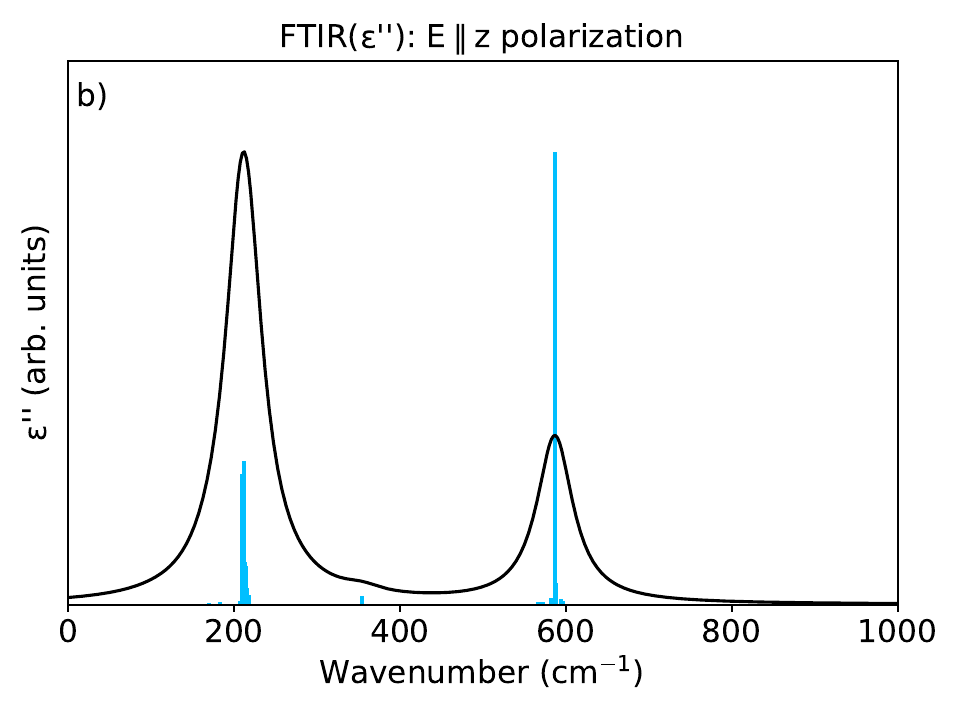}
    \caption{(a): Calculated Raman spectra in x(yz)$\overline{\mathrm{x}}$ polarization for three different SQS (labeled by the numbers 1-3) of LiNb$_{29.2}$Ta$_{70.8}$O$_3$. (Red) corresponds to the configuration which is used for all subsequent comparisons.
    (b): Imaginary part of the dielectric function of LNT calculated for light polarized along the extraordinary crystal axis (z-axis). The blue bars denote the contribution of single phonon modes to the complete spectrum in arbitrary units}
    \label{fig:sqs_comparison}
\end{figure}

LNT solid solutions are modeled with special quasi\-ran\-dom structures (SQS) created as 
described in \cite{WeiSQS,JoergSQS,VANDEWALLE201313}. In particular, 1$\times$1$\times$2 
repetitions of the orthorhombic unit cell are employed as supercells to model solid 
solutions containing a 70.8\,\% Ta-concentration. 4$\times$4$\times$1 Monkhorst-Pack K-point meshes are used, leading to a similar K-point density as for the unit cells. The cell parameters are determined by optimizing the lattice for different volumes and fitting the resulting energies to the Murnaghan equation of state \cite{Tyuterev2006}. The number of conduction bands used for the calculation of the electronic contribution to the dielectric function is chosen to obtain the same conduction/valence ratio as for the primitive cells when modeling the end compounds LNO and LTO.
Three different SQS are employed for the calculations presented in this work. The \textit{de facto} equivalence of these SQS is verified by comparing their calculated Raman signals. As seen in Fig.~\ref{fig:sqs_comparison}, the calculated spectra only differ marginally. This confirms, that the SQS well describe the same material, even though their atomistic distribution is different. Therefore, only one SQS is used as a reference throughout this work.
For the mixed crystals, the additional effect of phonons folding back onto the $\Gamma$-point has to be taken into account: Because a supercell is needed to describe the random alloy, there will be many more phonon modes present in their calculations. This will lead, in general, to more peaks in the spectra. Here, however, some of these folded modes lie so close to each other in their frequencies, that the applied Gaussian smearing will result in fewer, but broader peaks in the spectrum [see Fig.~\ref{fig:sqs_comparison} (b)]. The arguments regarding group theory, mentioned in Sec. II D, for the end compounds, strictly speaking, do not apply for the mixed crystals. Therefore, we denote the maxima instead with primed labels. For example, the frequency A$'_1$-TO$_1$ represents all modes that contribute to a single peak, which behaves comparable to the A$_1$-TO$_1$ in the end compounds. Even though these frequencies do not hold any information regarding a displacement pattern (except for their symmetry), their frequencies open up an interpretation regarding the Ta content of a given sample. Shifts in frequency depending on the Ta content have been explained in Ref.~\cite{RuesPRB}: For most modes, LNO features higher mode frequencies, which can be explained by the mass difference of niobium (92.906~a.u.) as compared to tantalum (180.948~a.u). Exceptions to this rule are modes involving mostly a movement of the oxygen cage (e.g. A$_1$-TO$_3$ and high frequency E-modes): Here, the shorter Ta-O bond length compared to Nb-O suggest a stronger bond, which results in higher frequencies in LTO for these modes.

\subsection{IR spectra}

In Raman or FTIR spectroscopy not only phonon frequencies, but also the scattering efficiencies or oscillator strength, respectively, are probed. For an accurate comparison of theoretical and experimental data, therefore, not only the exact frequencies but equally the relative intensities need to be calculated, which proved extremely valuable in the past in assignment of phonon features \cite{SimoRaman15,Hermet2007,Liebhaber2016,Plaickner2021,Neufeld2023}. 

(FT)IR spectroscopy directly probes the real and imaginary parts of the dielectric function $\varepsilon$ of a material in the respective phonon frequency range, i.e. the direct contribution of the vibrating lattice to the dielectric response.
The dipole moment $\vec{p}$ of a crystal is defined as:

\begin{equation}
\label{eq:dipole}
    \vec{p}=\sum\limits_b\left(\sum\limits_{i,a}Z_{iab}\vec{r}_i\hat{e}_a\right)\hat{e}_b\text{,}
\end{equation}

\noindent
where $\vec{r}_i$ denotes the position of ion $i$, and $Z_{iab}$ is its effective charge tensor 
(i.e. the Born charge). The vectors $\hat{e}_{a,b}$ label the cartesian unit vectors and 
the indices $a,b$ the cartesian coordinates.

The change of $\vec{p}$ with respect to a phonon with eigenmode $Q$ can be approximated 
by a symmetric difference quotient:

\begin{equation}
\label{eq:change}
    \frac{\partial\vec{p}}{\partial Q} \sim  \sum\limits_b\left(\sum\limits_{i,a}Z_{iab}\vec{Q}_i\hat{e}_a\right)\hat{e}_b\text{,}
\end{equation}

\noindent
where $\vec{Q}_i$ denotes the movement of ion $i$ within the displacement pattern $Q$. 
The imaginary part of the ionic contribution to the dielectric function can then be 
calculated as in Ref.~\cite{Gonze1997}:

\begin{align}
\label{eq:gonze}
    \Im{\left(\varepsilon^{\text{ion}}_{ab}(\omega_m)\right)}
   &\sim\left(\frac{\partial p_a}{\partial Q}\right)\left(\frac{\partial p_b}{\partial Q}\right)^{*}\nonumber
   \\
   &=\left(\sum\limits_{i,a'}Z_{ia'a}\vec{Q}_i\hat{e}_{a'}\right)\left(\sum\limits_{i,a'}Z_{ia'b}\vec{Q}_i\hat{e}_{a'}\right)^*\text{,}
\end{align}

\noindent
with $\omega_m$ the TO frequency of mode $m$ with displacement pattern $Q$. The 
contribution of all phonon modes to the IR spectra is obtained by calculating the 
sum over all modes $m$. In the harmonic approximation, these contributions will only 
yield delta peaks at the TO phonon frequencies $\omega_m$.
The real part of the ionic contribution to the dielectric function can be obtained as shown in Ref.~\cite{Gonze1997} as follows:

\begin{equation}
\label{eq:KK}
    \Re{\left(\varepsilon^{\text{ion}}_{ab}(\omega)\right)}
=\frac{4\pi}{V_0}\sum\limits_m\frac{\Im{\left(\varepsilon^{\text{ion}}_{ab}(\omega)\right)}}{\omega_m^2-\omega^2}\text{.}
\end{equation}


\subsection{Raman spectra}

In contrast to FTIR, Raman spectroscopy is an inelastic scattering technique, where a photon at much higher energies, typically in the visible range, exchanges energy with a phonon. Subsequently, a photon at a shifted frequency is emitted. This second-order effect is mediated via the electronic contribution of the dielectric response. The Raman tensor $\alpha$ is defined as the change of the polarizability with respect to a phononic eigenmode $Q$:

\begin{equation}
\label{eq:raman}
    \alpha_{m,ab}(\omega) = 
\frac{\partial \varepsilon^{\text{el}}_{ab}(\omega)}{\partial Q}
\sim\varepsilon_{\vec{R}_+,ab}^{\text{el}}(\omega)-\varepsilon_{\vec{R}_-,ab}^{\text{el}}(\omega)\text{,}
\end{equation}

\noindent
with $\varepsilon^{\text{el}}(\omega)$ being the electronic contribution to the dielectric 
function and $\omega$ the frequency of the incident laser light. Here, we calculate the
derivative evaluating $\varepsilon^{el}$ at two opposite, symmetric ionic displacements 
$\vec{R}_\pm$. The electronic contribution to the dielectric function is calculated within 
the independent particle approximation \cite{Gajdos2006}:

\begin{widetext}
\begin{equation}
\label{eq:lin}
    \Im[\varepsilon^{el}_{ab}(\omega)] \sim 
    \lim_{q\to 0}\frac{1}{q^2}\sum_{c,v,\vec{k}}2\delta(E_{c\vec{k}}-E_{v\vec{k}}-\omega) \times 
    \braket{u_{c\vec{k}+\vec{e}_aq}|u_{v\vec{k}}}\braket{u_{c\vec{k}+\vec{e}_bq}|u_{v\vec{k}}}^{*}\text{,}           
\end{equation}
\end{widetext}

\noindent
where $E_{c\vec{k}}$ and $E_{v\vec{k}}$ denote electronic energies at the conduction 
and valence band at k-vector $\vec{k}$ respectively and $\ket{u}$ denotes the 
corresponding Bloch function. Using Kramers-Kronig relations, we can obtain the real part of the electronic contribution to the dielectric function:

\begin{equation}
\label{eq:kk2}                                                                                                                            
    \Re[\varepsilon^{el}_{ab}(\omega)] = 
    1+\frac{2}{\pi}\int^{\infty}_{0}\frac{\Omega  \Im[\varepsilon^{el}_{ab}(\Omega)]}{\Omega^2-\omega^2+i\eta} \text{d}\Omega\text{,}
\end{equation}

\noindent
with $\eta$ being a small number to preserve the convergence of the integral.
The intensity of the Raman signal can be finally calculated as:

\begin{equation}
\label{eq:raman2}
    I_m(\omega)[s(ab)i] \sim 
    |\hat{e}_s\alpha_{m,ab}(\omega)\hat{e}_i|^2 \frac{(\omega-\omega_m)^4}{\omega_m}(n+1)\text{,}
\end{equation}

\noindent
where $\hat{e}_i$ and $\hat{e}_s$ denote the polarization of the incident and scattered 
light, respectively, and $n$ is the Bose-Einstein distribution.

In contrast to the theoretical data, the experimental spectra are subject to finite line width, due to thermal broadening, resolution limits, or crystal defects. This is not captured in our calculations. Therefore, artificial Gaussian smearings of \SI{5}{\per\centi\meter} width are applied to the calculated spectral intensities (both IR and Raman), to obtain spectra readily comparable to the measured data.


\subsection{LO-TO splitting}
In order to calculate LO-TO splitting of phonon frequencies at the $\Gamma$-point, a dipole-dipole interaction term is added to the dynamical matrix, which reads as \cite{Gonze1994, Gonze1997}:
\begin{equation}\label{eq:NAC}
    \frac{1}{\sqrt{m_jm_{j'}}}\frac{4\pi}{V_0}\frac{\left(\sum_cZ_{jca}\hat{q}_c\right)\left(\sum_cZ_{jcb}\hat{q}_c\right)}{\sum_{cc'}\hat{q}_c\varepsilon_{\infty,cc'}\hat{q}_{c'}}\text{,}
\end{equation}

\noindent
where $\hat{q}$ denotes a normalized reciprocal vector, $m_j$ is the mass of ion $j$, and $Z$ denotes the effective charge tensor of ion $j$. The letters $a$, $b$, $c$, and $c'$ denote cartesian indices. $\varepsilon_\infty$ is the electronic contribution of the static dielectric function. Since all these variables were already calculated in order to obtain IR- and Raman spectra, this comes at almost no additional computational cost.

\subsection{Phenomenological model}
The generalized Lyddane-Sachs-Teller (LST)
relation \cite{LST1,CHAVES1973865} can be employed to estimate the 
phononic contribution to the material’s optical response in an approximate manner. The generalized LST connects the high frequency permittivity $\varepsilon_\infty$ to the static permittivity $\varepsilon_0$ through

\begin{equation}
\label{eq:LST}
\varepsilon_0=
\varepsilon_\infty\prod_{j=1}^{n}
\frac{\omega^2_{\mathrm{LO}j}}{\omega^2_{\mathrm{TO}j}}\text{.}
\end{equation}

\noindent
This is very valuable, because the high-frequency 
dielectric function can be readily calculated from the electronic structure 
within density functional theory, but the calculation of the vibrational contributions in the limit of low frequencies - and therefore the static permittivity - is more challenging.

\section{\label{sec:res}Results}
\subsection{Experimental results}

\begin{figure*}
  \includegraphics[scale=0.2]{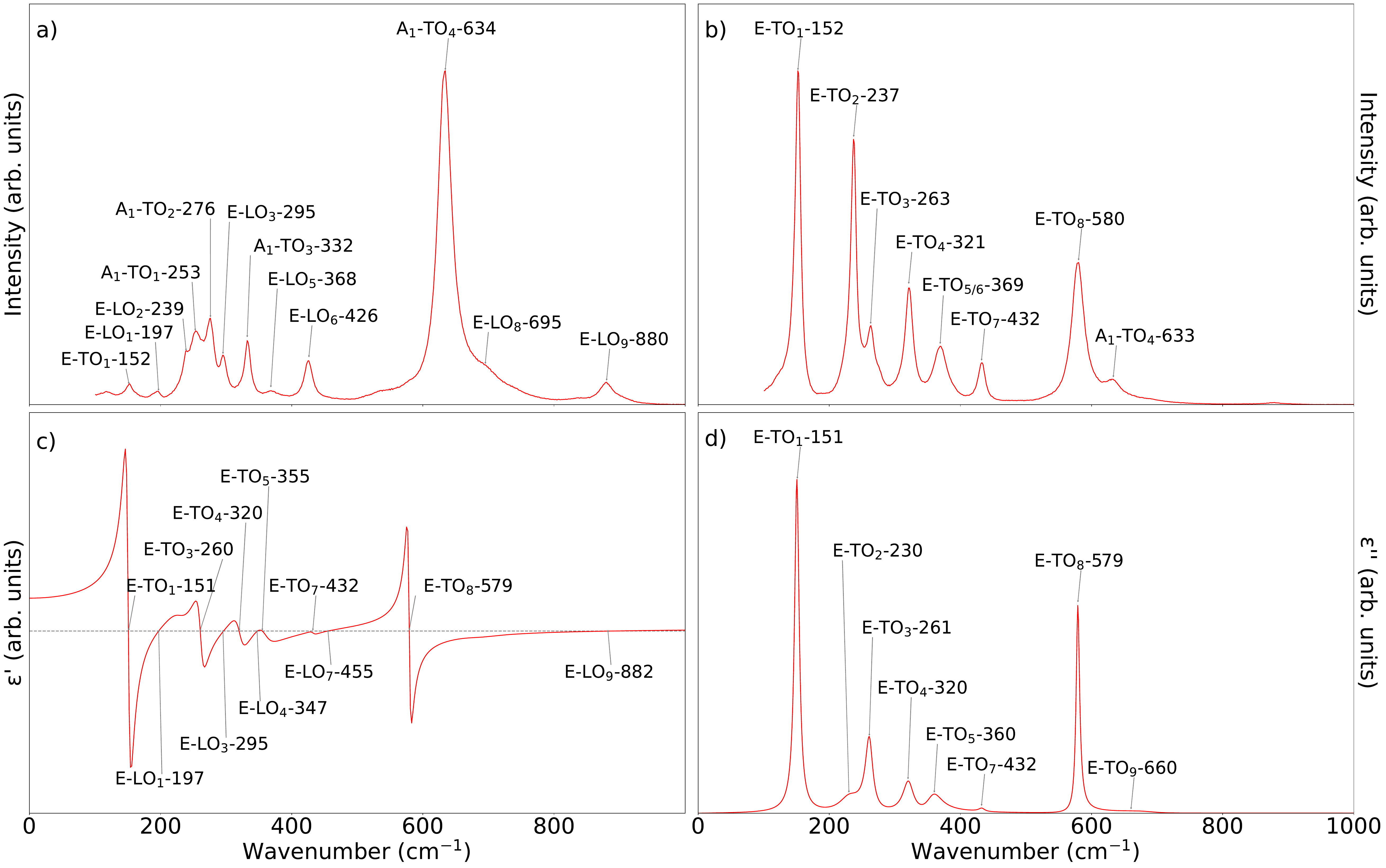}
  \caption{\label{fig:examplary}Top: a) Raman spectra of LNO for polarization y(xx)$\overline{\rm{y}}$ (left) and b) y(zx)$\overline{\rm{y}}$ (right); Bottom: c) Real and d) imaginary part of the dielectric function for light polarized along the crystallographic x direction, extracted from FTIR spectra. All spectra are normalized with respect to their highest peak. In the Raman spectra, the A$_1$-TO and E-LO modes (a), as well as the E-TO modes (b) can be assigned by applying the symmetry considerations from Table \ref{tab:porto}. The E-LO modes can be extracted as zero crossings in $\varepsilon'$. Additionally, the E-TO modes can be extracted from peak positions in $\varepsilon''$ (d). The redundancies in these four graphics serve as an appraisal for the measurements and calculations.}
\end{figure*}

To unambiguously identify and assign all phonon modes in the LNT material family the measured experimental data, for both, Raman and FTIR spectroscopy were systematically analyzed and compared, while considering the selection rules listed in Tables \ref{tab:porto} and \ref{tab:FTIR}. To illustrate both, the process of comparing different Raman spectra, and the comparison of Raman and FTIR data, Figure~\ref{fig:examplary} shows several datasets obtained on the LNO samples. All spectra shown contain signatures of E-type phonons, including both LO and TO modes. In detail, subfigure a) displays a Raman spectrum in y(x,x)$\overline{\rm{y}}$ configuration, which according to the selection rules shows E-LO and A$_1$-TO modes, while b) shows a Raman spectrum in the crossed configuration y(z,x)$\overline{\rm{y}}$, which only addresses E-TO phonons. These Raman spectra are compared with c), the real part $\varepsilon'$ and d), the imaginary part $\varepsilon''$ of the ordinary-polarized dielectric function extracted from FTIR spectroscopy using the fitting process shown in Fig.~\ref{fig:FTIR_raw}.

For the dielectric function, not just the fitting as discussed above (Sec.~IIC) yields the phonon frequencies, but the phonon frequencies can also be visually assigned to specific features in the real and imaginary parts of the dielectric function, which can also be visually compared to the Raman results. Specifically, in the real part $\varepsilon'$ the TO-frequencies are indicated by the zero-crossing of the dielectric function at the central wavelength of each oscillator (zero-crossing at a downward slope), while the zero-crossing on the rising flanks of the dielectric function indicate the associated LO phonon frequency. These two boundaries define the Reststrahlen band in the reflectivity spectrum for each phonon mode visible in the raw spectrum in Fig.~\ref{fig:FTIR_raw}. The positions of the falling (rising) flank zero crossings can be readily compared with the TO- (LO-) phonon peaks in the Raman spectra. Please note, the zero-crossing (up or down) of \emph{each} oscillator highlighted in the graph does not necessarily indicate a zero-crossing of the \emph{total} dielectric function. 

In contrast to the real part, the imaginary part of the dielectric function resembles in its structure an emission-type spectrum with peaks centered at the TO-frequencies. This structure looks very similar to the Raman spectrum of TO-phonons, which therefore allows a direct visual comparison. Please note, that while the spectra have a similar shape, the spectra originate from different physical processes, i.e. a phonon with a strong absorption in the dielectric function does not necessarily mean a high scattering efficiency in Raman scattering.

In the cross-polarized Raman spectrum in Fig.~\ref{fig:examplary} b) eight peaks can be identified. Out of these, seven belong to E-TO phonons, while the peak at 633\,cm$^{-1}$ belongs to the A$_1$-TO$_4$ that is not fully suppressed by the analyzer. In detail, the visible peaks belong to the E-TO$_1$, -TO$_2$, -TO$_3$, -TO$_4$, -TO$_{5/6}$, -TO$_7$, and -TO$_8$ modes. Similarly, the imaginary dielectric function shows six relatively strong peaks (E-TO$_1$, -TO$_2$, -TO$_3$, -TO$_4$,  -TO$_5$, -TO$_8$) and two very weak, but identifiable peaks (E-TO$_7$, -TO$_9$). Here, it can be seen that the E-TO$_9$ at 660\,cm$^{-1}$ has a very weak intensity in both Raman and FTIR, which was previously observed as well \cite{Margueron2012,SimoRaman15,RuesPRB} and is also found in the DFPT calculations (Fig. \ref{fig:LNO_x_sup} in the supplement). Its behavior is different from its behavior in LTO, where the E-TO$_9$ has a much stronger scattering efficiency in Raman spectroscopy, but is still very weak in FTIR spectroscopy (see Fig. \ref{fig:LTO_x_sup} in the supplement). Due to its low scattering efficiency, the E-TO$_9$ mode was often assigned ambiguously in the past \cite{Margueron2012,Hermet2007,RuesPRB, SimoRaman15}. Further, it can be noticed that the E-TO$_5$ and E-TO$_6$ frequencies are very close (360\,cm$^{-1}$ vs. 367\,cm$^{-1}$). While the TO$_6$ is dominating in Raman scattering, the TO$_5$ is the dominating feature for the FTIR spectrum. Previous work has indicated that the scattering efficiency and their behavior in both phonons are similar for both, LTO and mixed crystals. As a result, the Raman and FTIR spectra were only fitted with one peak representing the TO$_5$ or TO$_6$ mode, respectively. Only low temperature measurements of stoichiometric crystals allow to unambiguously differentiate these modes in Raman spectroscopy \cite{Margueron2012}. The origin in the different scattering efficiencies for both phonons and in both methods are also backed up by the DFPT calculations discussed below. Overall, the E-TO phonon frequencies determined from different Raman spectra, as well as by comparing FTIR and Raman spectra, match within $\pm$ 5\,cm$^{-1}$, which is in agreement with the combined resolution limits of both methods.

\begin{figure*}
    \includegraphics[scale=0.2]{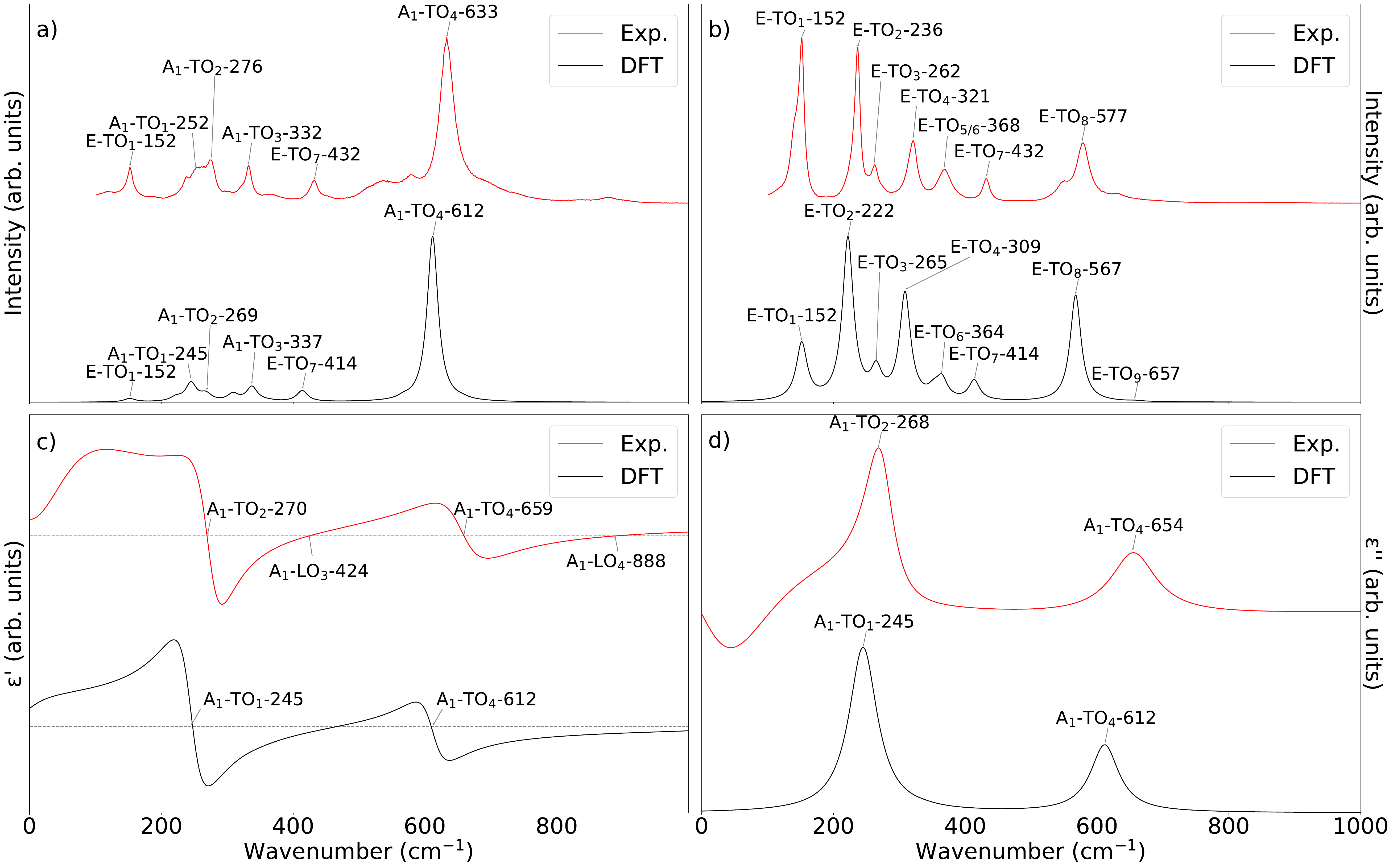}
    \caption{\label{fig:LNO_comparison_A}Top: a) Raman spectra of LNO for polarization x(yy)$\overline{\rm{x}}$ (left) and b) x(zy)$\overline{\rm{x}}$ (right); Bottom: c) Real and d) imaginary part of the dielectric function for light polarized along the crystallographic z direction, extracted from FTIR spectra. All spectra are normalized with respect to their highest peak. In the Raman spectra, the A$_1$-TO and E-TO modes can be unambiguously assigned, by applying the symmetry considerations from Table \ref{tab:porto}. The A$_1$-LO modes can be extracted as zero crossings in $\varepsilon'$. The redundancies in these four graphics serve as an appraisal for the measurements and calculations.}
\end{figure*}

The spectrum in Fig.~\ref{fig:examplary} a) shows a Raman spectrum in y(x,x)$\overline{\rm{y}}$ configuration, which shows both E-LO and A$_1$-TO phonons. The A$_1$-TO phonons are dominating in intensity. Here, they have a strong scattering efficiency leading to strong overlaps in the 200 to 400\,cm$^{-1}$ region. The A$_1$-TO phonon peaks can be readily excluded by comparison with spectra in x(z,z)$\overline{\rm{x}}$ [Fig.~\ref{fig:LNO_z_sup} (a)] or x(y,y)$\overline{\rm{x}}$ configurations [e.g. Fig.~\ref{fig:LNO_x_sup} (a)]. The remaining peaks, therefore, are exclusively E-LO phonons. E-LO phonons are only accessible in this single scattering configuration for back scattering, which in the past lead to the most ambiguity in assignment for this phonon branch \cite{Hermet2007,Margueron2012}. Therefore, the comparison to the real part of the dielectric function (and to the calculations) is particular helpful for a correct assignment. Again, by comparison, a good agreement is found and all nine E-LO phonons can be identified and a frequency can be determined.

Similar to the discussion above, the Raman and FTIR data for all samples, LNO, LTO, and LNT, was evaluated for different cuts. Additional plots and tabular summaries of all phonon frequencies are available in the supplemental file. These tables denote the mean phonon frequencies, determined by analysing all our Raman and FTIR spectra for all available crystal cuts. In the next step, the experimental data will be compared exemplarily with the calculations from DFPT.

\subsection{Comparison between theory and experiment}

\begin{figure*}
    \includegraphics[scale=0.2]{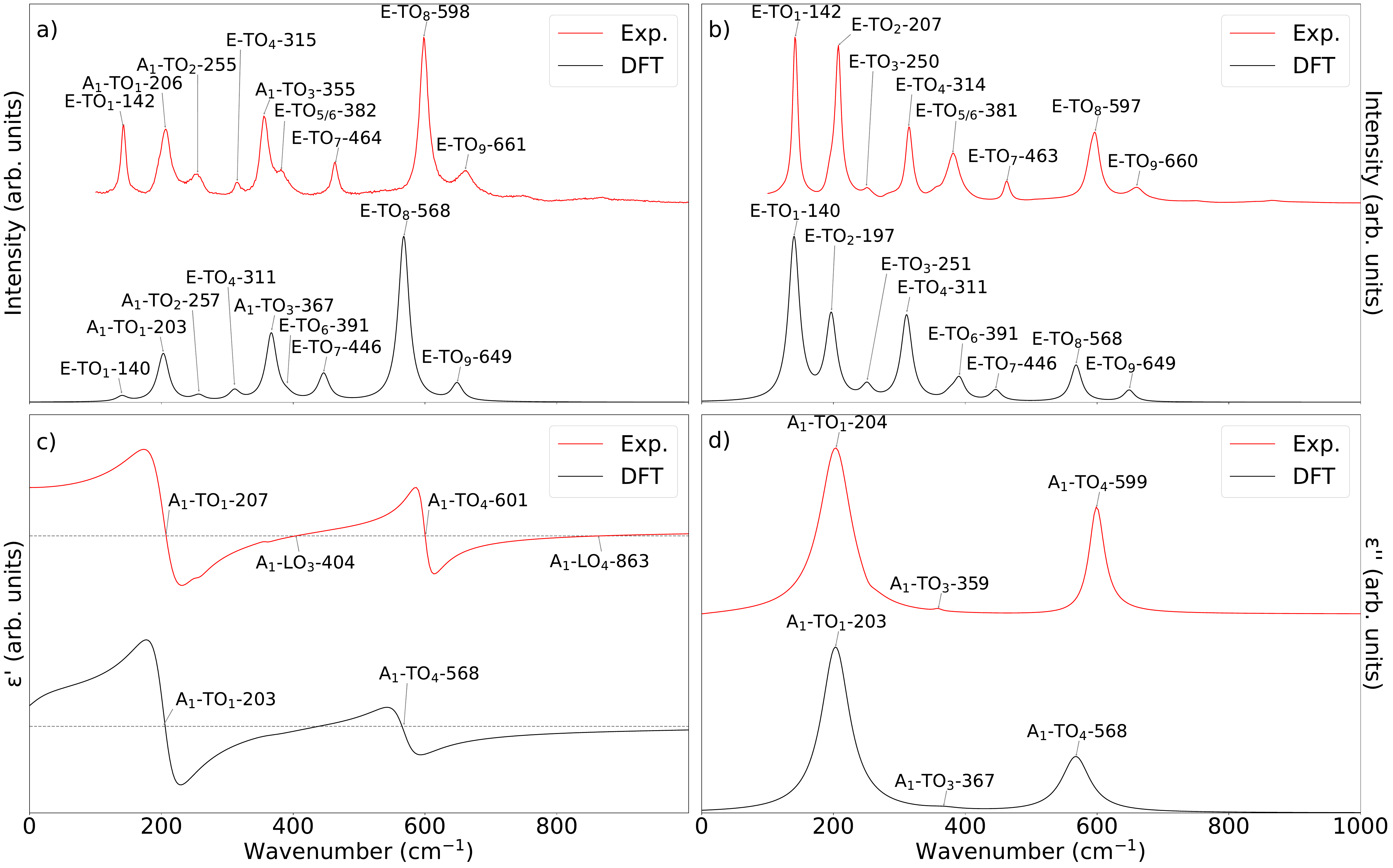}
    \caption{\label{fig:LTO_comparison_A}Top: a) Raman spectra of LTO for polarization x(yy)$\overline{\rm{x}}$ (left) and b) x(zy)$\overline{\rm{x}}$ (right); Bottom: c) Real and d) imaginary dielectric functions for light polarized along the crystallographic z direction, extracted from FTIR spectra. All spectra are normalized with respect to their highest peak. In the Raman spectra, the A$_1$-TO and E-TO modes can be unambiguously assigned, by applying the symmetry considerations from Table \ref{tab:porto}. The A$_1$-LO modes can be extracted as zero crossings or poles in $\epsilon'$. The redundancies in these four graphics serve as an appraisal for the measurements and calculations.}
\end{figure*}

\begin{figure*}
    \includegraphics[scale=0.2]{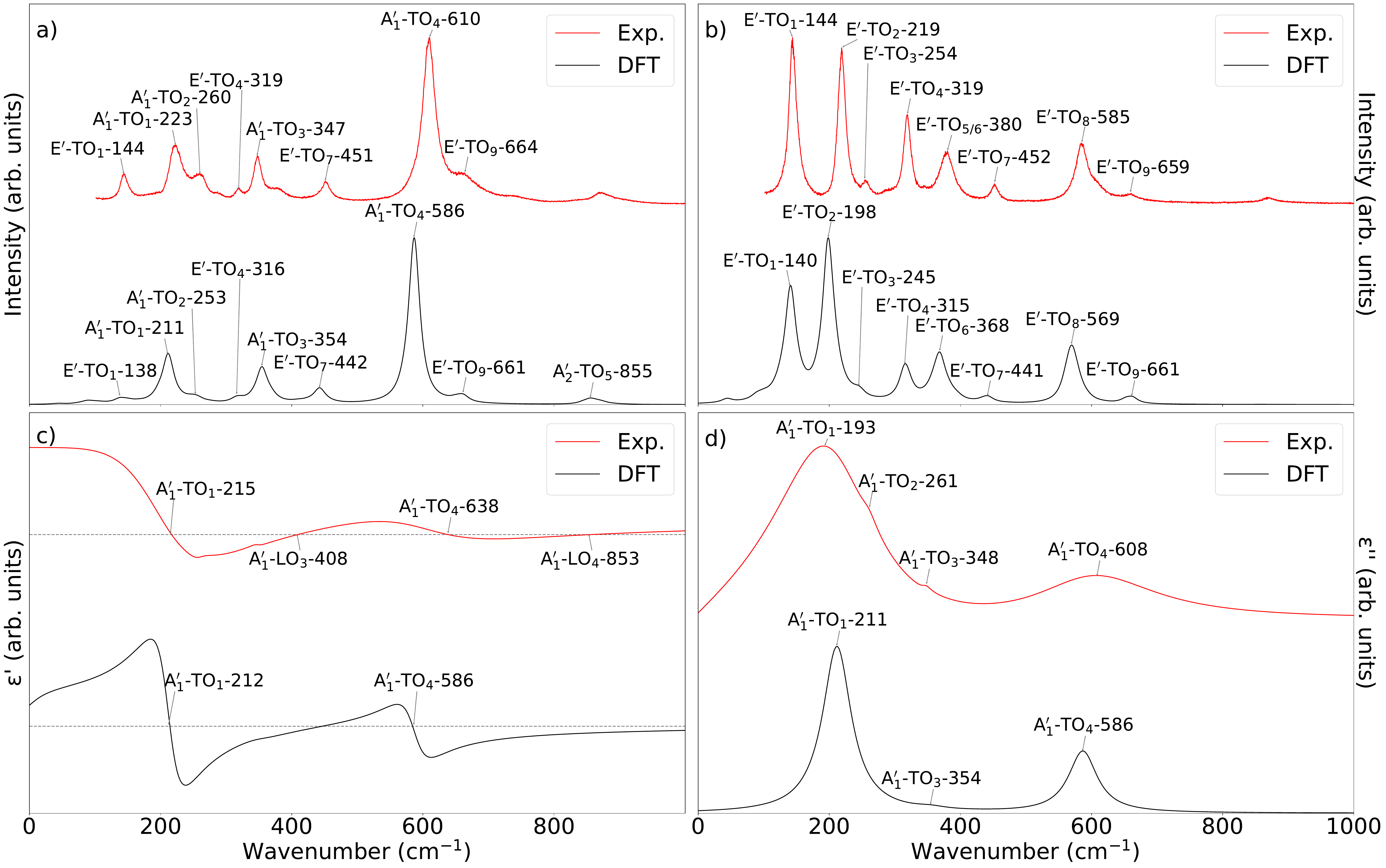}
    \caption{\label{fig:LNT_comparison_A}Top: a) Raman spectra of LNT for polarization x(yy)$\overline{\rm{x}}$ (left) and b) x(zy)$\overline{\rm{x}}$ (right); Bottom: c) Real and d) imaginary dielectric functions for light polarized along the crystallographic z direction, extracted from FTIR spectra. All spectra are normalized with respect to their highest peak. In the Raman spectra, the A$'_1$-TO and E$'$-TO modes can be unambiguously assigned, by applying the symmetry considerations from Table \ref{tab:porto}. The A$'_1$-LO modes can be extracted as zero crossings or poles in $\varepsilon'$. The redundancies in these four graphics serve as an appraisal for the measurements and calculations.}
\end{figure*}

The calculated and measured TO phonon frequencies of LNO and LTO are given in Tables \ref{tab:modes_LN}, and \ref{tab:modes_LT} (in the supplement), respectively. As the A$_2$ modes are Raman and IR silent for ideal crystals, we just report the calculated frequencies, although - as seen below - we may see indications of those phonons activated in the measured spectra due to defects or stoichiometry. A detailed discussion of the difference in frequencies for the same modes in LNO and LTO is given in previous work \cite{RuesPRB} and will not be repeated here. Here, we will specifically focus on the calculated spectra for both, the dielectric function and Raman signal. Because we only calculated Raman and FTIR spectra for TO frequencies, we limit our comparison of Raman spectra to the x($\cdot\cdot$)$\overline{\mathrm{x}}$ configuration.

Overall, the calculated and measured TO phonon frequencies are all in good agreement with each other and lay mostly within the accuracy of both experimental setups, as well as with available previous computational and experimental results \cite{SimoRaman15,Margueron2012,Hermet2007}. The largest deviation for LNO occurs for the E-TO$_2$ mode with a discrepancy of around 6\% (\SI{14}{\per\centi\meter}) [see e.g. Fig. \ref{fig:LNO_comparison_A} (a) or Tab. \ref{tab:modes_LN}]. The overall small deviations in frequency between experiment and theory is mostly caused by using fixed lattice constants in the simulations, which are in good agreement to experimental values. 

The dielectric functions show the same number of peaks/features and similar intensities in the real and imaginary part for both investigated polarization directions for LNO, LTO and LNT. We note, that the simulated $\varepsilon'$ does not allow for an assignment of the LO modes, as no information from the simulation leading to a LO-TO splitting is provided at this point. The peaks and shape of $\varepsilon'$ and $\varepsilon''$ do, however, coincide to the experimental data in all cases. Please note, the peak width in the calculated data is arbitrarily introduced and knowledge of the damping constants from theory is not available.

\begin{figure*}[h]
  \includegraphics[scale=0.2]{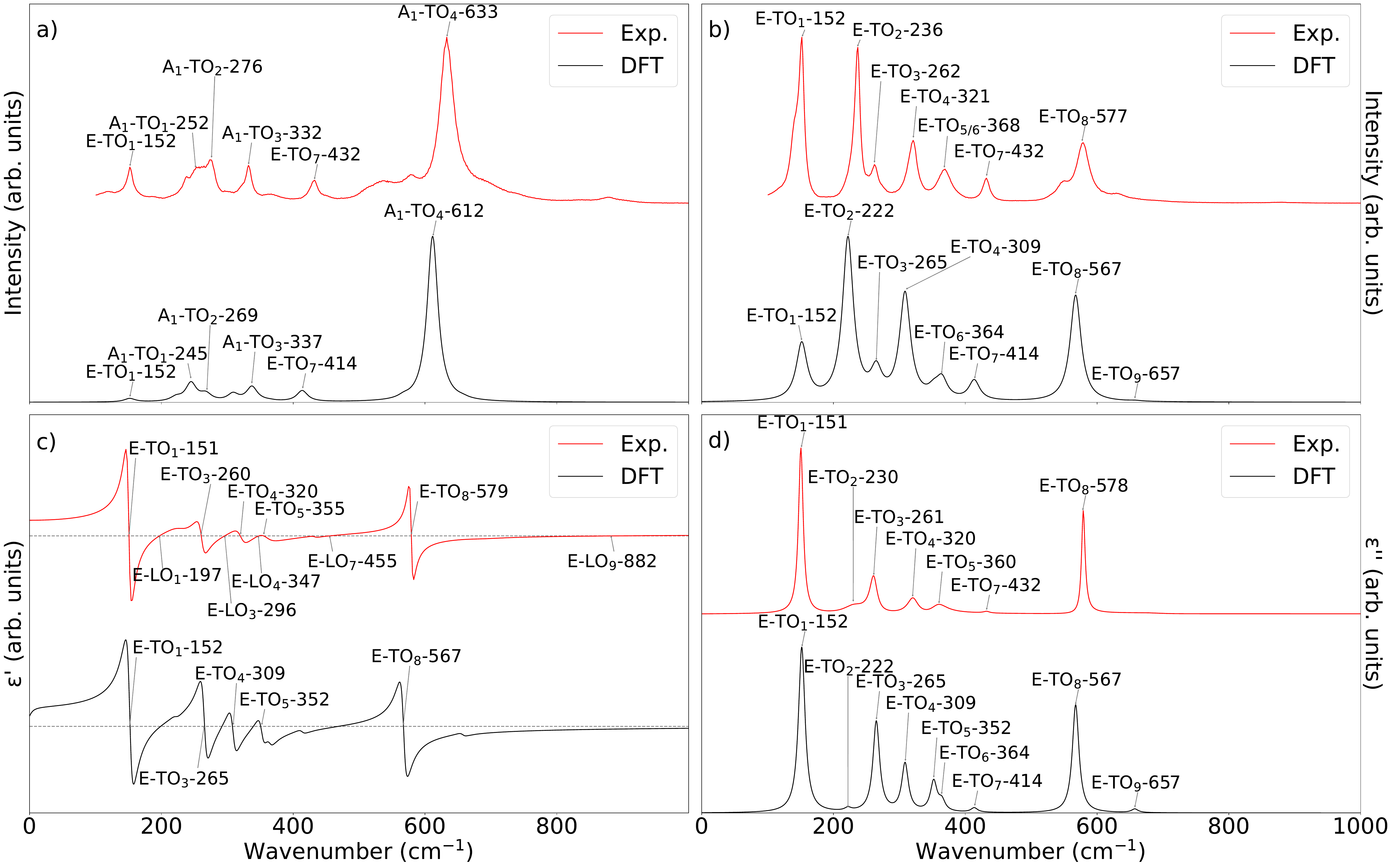}
  \caption{\label{fig:LNO_x_sup}Top: a) Raman spectra of LNO for polarization x(yy)$\overline{\rm{x}}$ (left) and b) x(zy)$\overline{\rm{x}}$ (right); Bottom: c) Real and d) imaginary dielectric functions for light polarized along the crystallographic x direction, extracted from FTIR spectra. All spectra are normalized with respect to their highest peak. In the Raman spectra, the A$_1$-TO and E-TO modes can be unambiguously assigned, by applying the symmetry considerations from Table \ref{tab:porto}. The E-LO modes can be extracted as zero crossings or poles in $\varepsilon'$. The redundancies in these four graphics serve as an appraisal for the measurements and calculations.}
\end{figure*}

\begin{figure*}
  \includegraphics[scale=0.2]{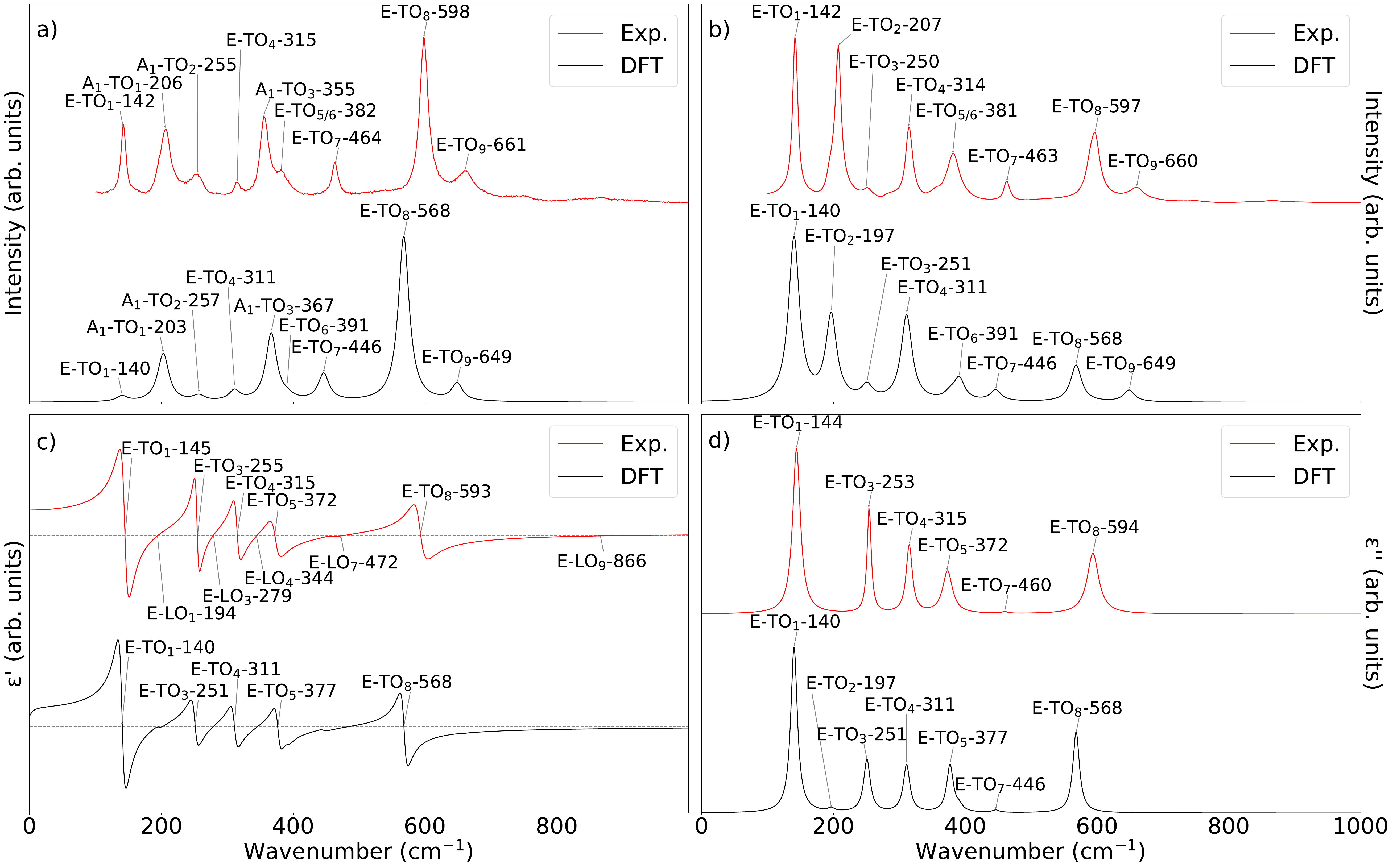}
  \caption{\label{fig:LTO_x_sup}Top: a) Raman spectra of LTO for polarization x(yy)$\overline{\rm{x}}$ (left) and b) x(yz)$\overline{\rm{x}}$ (right); Bottom: c) Real and d) imaginary dielectric functions for light polarized along the crystallographic z direction, extracted from FTIR spectra. All spectra are normalized with respect to their highest peak. In the Raman spectra, the A$_1$-TO and E-TO modes can be unambiguously assigned, by applying the symmetry considerations from Table \ref{tab:porto}. The E-LO modes can be extracted as zero crossings or poles in $\varepsilon'$. The redundancies in these four graphics serve as an appraisal for the measurements and calculations. Note, that the E$_2$-TO-207 mode is not visible in the measured FTIR spectra, but pronounced in Raman. In the calculations this mode can be observed for both, FTIR and Raman.}
\end{figure*}

The splitting as well as the different Raman and IR activity of the E-TO$_5$ and E-TO$_6$ modes can be easily verified by our calculations [e.g. Fig. \ref{fig:LNO_x_sup} for LNO, and Fig. \ref{fig:LTO_x_sup} for LTO]: Here the E-TO$_5$ and E-TO$_6$ mode (352\,cm$^{-1}$ vs 364\,cm$^{-1}$ for LNO) can be clearly separated in the spectra. As observed in the experiment, the E-TO$_6$ is almost IR inactive (invisible in the spectra), but has a higher Raman activity than E-TO$_5$, while the inverse statement is true for the E-TO$_6$.

The measured dielectric function for E $\parallel$ z  shows far broader peaks than for E $\parallel$ x in LNO (comparing Fig.~\ref{fig:LNO_comparison_A} and Fig.~\ref{fig:examplary}). The phonon linewidth associated with this broadening is typically dependent on the stoichiometry of the sample, its homogeneity and the presence of defects, and the (natural) lifetime of the phonon itself. The determination of phonon lifetimes from first principles requires the calculation of phonon-phonon interations, beyond the harmonic approximation considered here. Such calculations by Fu et al. suggest no significant longer lifetimes for these modes, compared to the E-modes \cite{D1CE01323H}. We therefore conclude, that our samples are more homogeneous in the x-direction, but include inhomogeneities in z-direction, which agrees well with the nature of many polar defects and the nature of the ferroelectric domains that align all parallel to the z-axis. For better comparison, we increase the Gaussian smearing applied to the calculated spectra to \SI{25}{\per\centi\meter} for this direction, which will lead to broader peaks in e.g. Fig.~\ref{fig:LNO_z_sup} compared to e.g. Fig.~\ref{fig:LNO_x_sup} in the supplement. It is well known that the stoichiometry, in particular the lithium deficiency, of LNO and LTO crystals has a significant impact on their optical properties \cite{Kovacs1997, Debnath2014}. Nevertheless, we cannot determine a significant difference between the calculated and measured spectra for E $\parallel$ z in neither the peaks position, nor their intensity. The broadening of the spectra is not as pronounced for our LTO crystals, hence, we suspect these samples to be relatively homogeneous, i.e. mostly defect free with few domains and domain walls and a homogeneous Li-vacancy distribution. Note, that more narrow spectra in LTO were observed as well in the past \cite{RuesPRB}.

The previously discussed effect of phonon folding makes an interpretation regarding the stoichiometry of the LNT crystal more challenging than for the end compounds (Note, that in an alloy, the distribution of Ta and Nb atoms is expected to be random). Therefore, each lattice site which breaks periodicity could be interpreted as an inhomogeneity. This means, we expect an overall broadening of the peaks (broader than in the end compounds), regardless of other defects, e.g. Li-inhomogenities. The A$'_1$-TO$_4$-like mode at \SI{587}{\per\centi\meter} has by far the highest IR activity. However, the peak resulting from the overlap of multiple modes at 210-\SI{215}{\per\centi\meter} results in an overall higher signal. Only a single mode at \SI{354}{\per\centi\meter} contributes to the shoulder at the same frequency. Therefore, in general, in the mixed crystals we cannot assign a peak in either the FTIR or Raman signal to a true single phonon eigenmode, but instead assign such a sum feature with a single frequency to its closest analogue in the end compounds and label it with a prime. 

Curiously, apart from the expected A$_1$ and E-type-like phonon modes, we note a significant peak in the calculated and measured Raman spectra for x(yy)$\overline{\rm{x}}$ polarization (Fig.~\ref{fig:LNT_comparison_A}) at 855\,cm$^{-1}$: In the calculations, this peak belongs to the A$'_2$-TO$_5$ mode, which is due to numerical limitations (i.e. in the dielectric function and the slightly imperfect eigenvector) not completely suppressed. However, this peak is also visible in the measured spectrum for LNT, and even for LNO and LTO in the same Raman configuration, but it is much less pronounced for the end compounds. Here, this peak could also correspond to a not completely suppressed E-LO$_9$ or A$_1$-LO$_4$ mode. However, its clear presence in the calculated spectra of the solid solution might also indicate that defects in the end compounds or the random Nb-Ta distribution in the mixed crystals can relax the selection rules, which enables the A$_2$-TO$_5$ mode to become Raman active. 

\begin{figure}
    \centering
     \hspace{-1.5cm}A$_1$-TO$_4$\hspace{1.5cm}E-TO$_9$
     \\
    \includegraphics[scale=0.2]{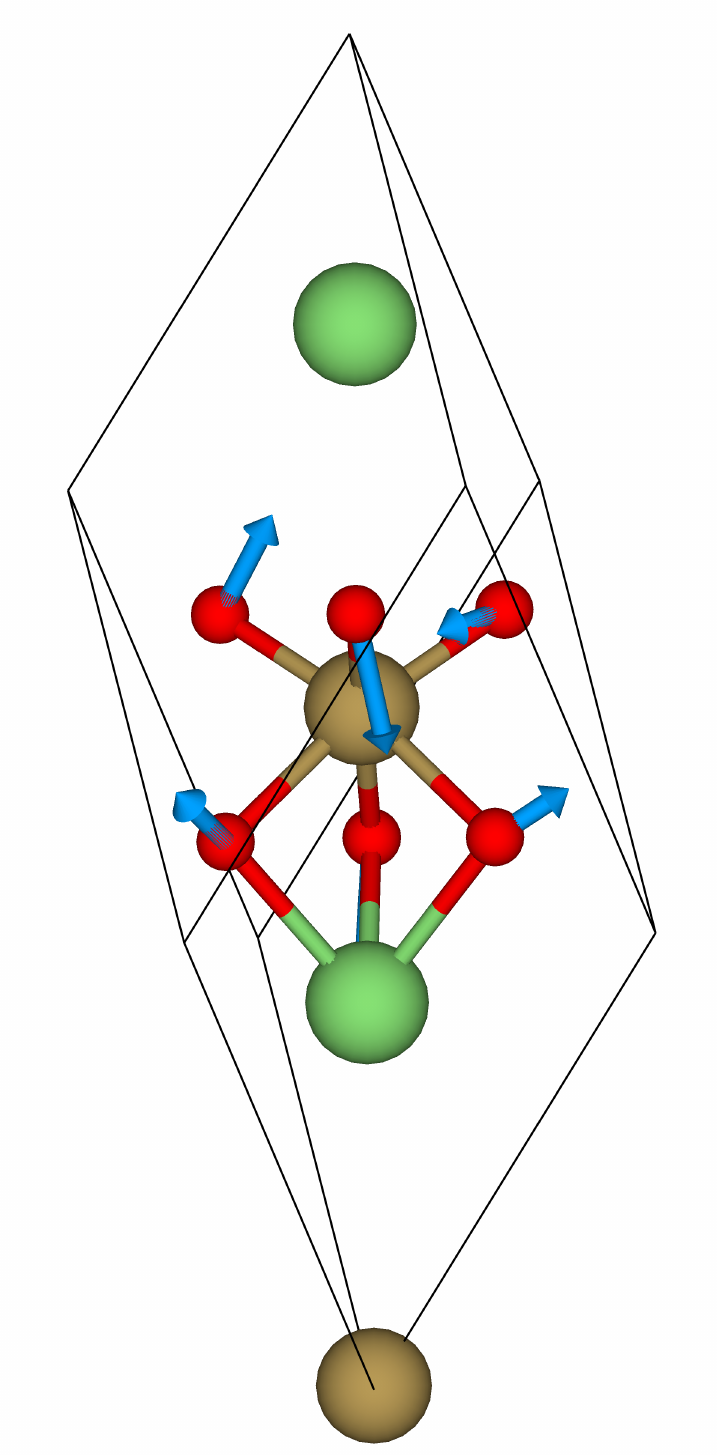}
    \includegraphics[scale=0.2]{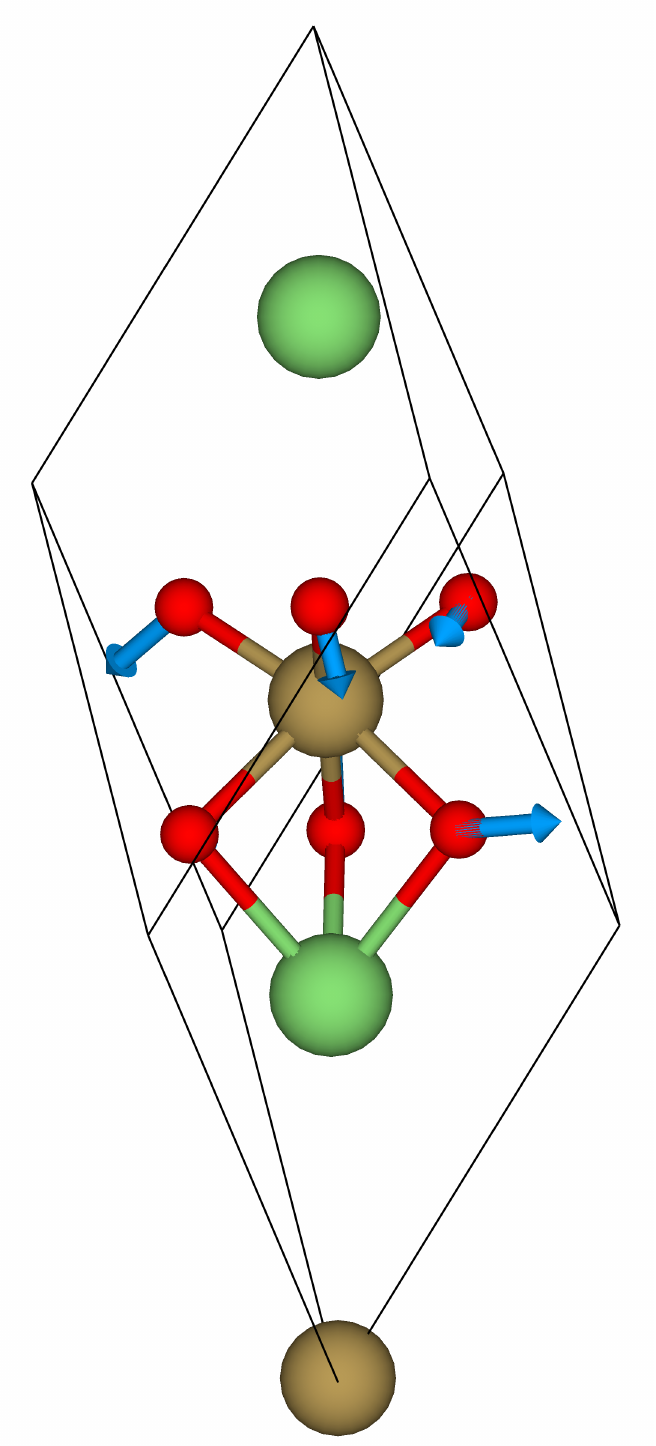}
    \includegraphics[scale=0.2]{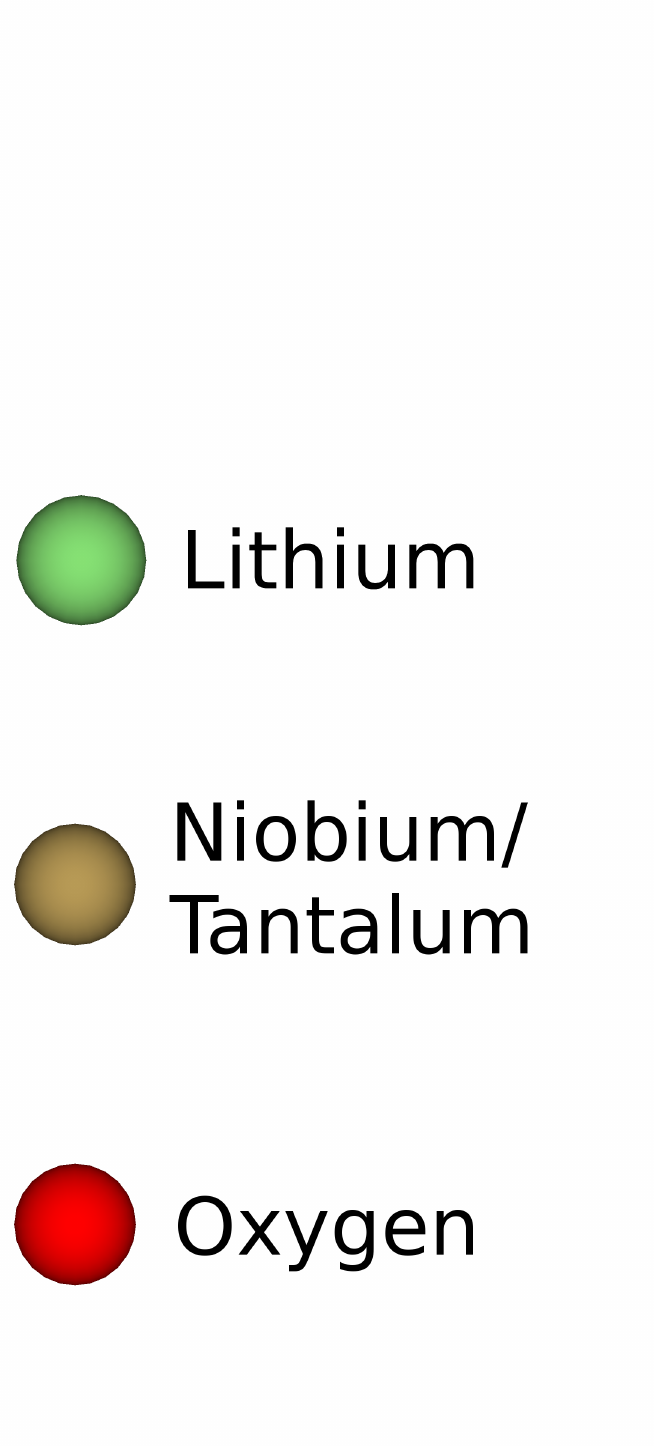}
    \caption{\label{fig:eigenmodes}Left: A$_1$-TO$_4$ eigenmode. Right: E-TO$_9$ eigenmode. The niobium/tantalum, lithium and oxygen ions are colored gold, green and red, respectively. The displacements are denoted by blue arrows.}
\end{figure}

For our simulations, the LO-mode frequencies cannot be directly extracted from the dielectric function. Instead, we apply Eq.~\eqref{eq:NAC} for different q-directions and track how the added dipole-dipole interaction affects the mode frequencies at $\Gamma$. Our results can be found in Tables \ref{tab:LO_modes_LN}, \ref{tab:LO_modes_LT} and \ref{tab:modes_LNT} in the supplement. Note, that we do not show A$_2$ modes in these tables, as they do not show any LO-TO splitting. Overall, the calculated and measured frequencies agree within an error of $9\%$, where the measured frequencies are for almost all modes slightly larger. This discrepancy can be easily explained by taking a closer look at Eq.~\eqref{eq:NAC}: Because of the underestimation of the electronic band gap in the DFT calculations (e.g. measured for LNO \SI{3.7}{\electronvolt} \cite{Dhar1990} vs. our DFT calculation with \SI{3.41}{\electronvolt}), a red-shift of the imaginary part of the dielectric function is introduced. This in turn gives rise to a larger value for $\varepsilon_\infty$, as Kramers-Kronig relations are used to calculate the real part. Thus, the added dipole-dipole interaction term is likely underestimated, which in conclusion leads to smaller LO-frequencies.
Especially large LO-TO splits can be observed for the high frequency A$_1$- and E modes (A$_1$-TO$_4$ - A$_1$-LO$_4$ and E-TO$_9$ - E-LO$_9$): Both of these modes involve a stretching of the oxygen cage, as well as slight movement of the niobium/tantalum within it (see Fig.~\ref{fig:eigenmodes}). Because this central ion carries a large effective charge, these movements introduce a strong dipole-dipole interaction, which in turn leads to a large LO-TO splitting.

\subsection{Calculation of $\varepsilon_0$}

\begin{figure}
    \includegraphics[scale=0.25]{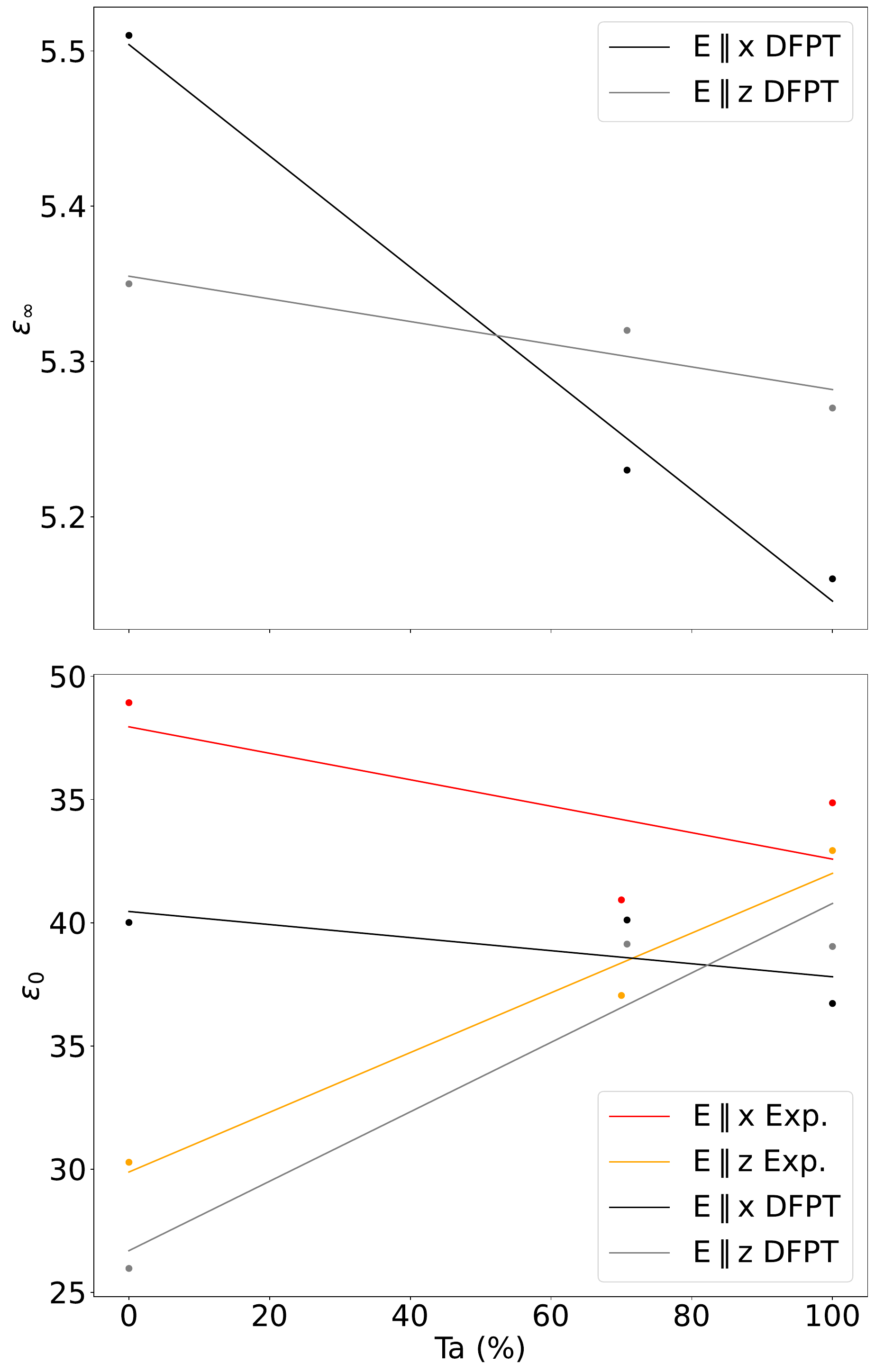}
    \caption{\label{fig:LST}$\varepsilon_\infty$ (top) and $\varepsilon_0$ (bottom) as a function of tantalum content of the LNT crystal family. The values for $\varepsilon_0$ are obtained using the LST relation [Eq.~\eqref{eq:LST}], whereas $\varepsilon_\infty$ is obtained by DFPT calculations. Only polarizations E $\parallel$ x and E $\parallel$ z are considered. For calculating $\varepsilon_0$, either the measured phonon frequencies (Exp., red and orange), or the calculated ones (DFPT, black and gray) are used.}
\end{figure}

Finally, we use the LST relation to estimate $\varepsilon_0$: From the DFPT calculation, we can extract $\varepsilon_\infty$ as the value of $\varepsilon'$ at excitation frequency zero. The extracted values for LNO are $\varepsilon_\infty^x=5.51$ and $\varepsilon_\infty^z=5.35$.
Inserting the calculated values for $\varepsilon_\infty$ and phonon frequencies for LNO into Eq.~\eqref{eq:LST}, we find $\varepsilon_0^x=40.0$ and $\varepsilon_0^z=26.0$. Both values are in good agreement to literature \cite{Barker1970, Fujii2006, Margueron2012, RuesPRB}. Using the measured phonon frequencies, we instead obtain $\varepsilon_0^x=48.9$ and $\varepsilon_0^z=30.3$, which is slightly larger, however also within the range typically reported in literature \cite{Barker1970, Fujii2006, Margueron2012, RuesPRB}. Repeating the calculations for LTO and LNT, we obtain $\varepsilon_0$ as a function of Ta content (see Fig.~\ref{fig:LST}). All extracted values, including LTO and LNT, can be found in Table \ref{tab:epsilon_0}. Note, that the overestimation of $\varepsilon_\infty$ leads to slightly higher values of $\varepsilon_0$, compared to the literature. In total, we thus suspect our results to be slightly overestimating $\varepsilon_0$. The uncertainty of $\pm$4\,cm$^{-1}$ in the measured frequencies results in an error of around 7\% for our obtained values for $\varepsilon_0$.

\begin{table}
\caption{\label{tab:epsilon_0} With Eq. \eqref{eq:LST} calculated values for $\varepsilon_0$, using either the calculated frequencies (DFPT) or the measured ones (DFPT+Exp.). Only polarizations E $\parallel$ x (z-cut) and E $\parallel$ z (x-cut) are considered. Available data \cite{Barker1970, Fujii2006, Margueron2012, RuesPRB} (Ref.) are reported for comparison.}
\begin{ruledtabular}
\begin{tabular}{ll|ll|lll}
& & \multicolumn{2}{c|}{$\varepsilon_\infty$} & \multicolumn{3}{c}{$\varepsilon_0$} \\
crystal & cut & DFPT & Ref. & DPFT & DFPT+Exp. & Ref. \\
\hline
\multirow{2}{*}{LNO} & z & 5.51 & 5.00 & 40.0 & 48.9 & 39.2-42.5 \\
& x & 5.35 & 4.60 & 26.0 & 30.3 & 23.6-26 \\
\multirow{2}{*}{LTO}& z & 5.16 & 4.50 & 36.7 & 44.9 & 34.7-42 \\
& x & 5.27 & 4.53 & 40.0 & 42.9 & 35.8-40 \\
\multirow{2}{*}{LNT}& z & 5.23 & 4.44 & 40.1 & 40.9 & 36.4 \\
& x & 5.32 & 4.39 & 36.7 & 37.0 & 32.1 \\

\end{tabular}
\end{ruledtabular}
\end{table}

\section{Summary}
In this work we have studied the vibrational properties of the lattice in the lithium niobate tantalate material family by studying the end compounds LTO and LNO, as well as an example of a mixed crystal LNT. To achieve a high redundancy in the analysis, we have systematically compared \emph{measured} Raman spectra and FTIR reflectance spectra, which allows us to extract the real and imaginary parts of the dielectric function, with \emph{calculated} Raman spectra and dielectric functions obtained from first principles. When applying group symmetry considerations and comparing all resulting spectra (experiment and theory), we can unambiguously assign all transverse optical and longitudinal optical phonon modes at the Brillouin zone center ($\Gamma$-point). In particular, by combining both, theory and experiment, the similarities and differences in the spectra, as well as their changes within mixed crystals, can be readily explained and interpreted. An example is the different intensity of quasi-silent modes in Raman or FTIR spectra, or even the appearance of symmetry forbidden modes like the A$_2$-TO$_5$ in Raman spectra. Here, the experiments and the theoretical calculations are in excellent agreement for all studied samples, demonstrating the high redundancy of this combined approach. Both, the theory and experimental data additionally allow us to obtain the phononic contribution to the dielectric function in the full LNT material family by applying the LST. Again, the calculated data are in excellent agreement to previous studies \cite{Barker1970, Fujii2006, Margueron2012, RuesPRB}. In principle, this methodology is not limited to the LNO-LTO crystal system, but this tested approach can be readily expanded to other crystals, doped or strained systems or domain wall spectra \cite{Ruesing2018,Rix22,Nataf2018}, where it is essential to understand and interpret the underlying physical effects and crystal structure by providing a quantitative interpretation of Raman or FTIR spectra.

\section*{Acknowledgements}
We gratefully acknowledge ﬁnancial support by the Deutsche Forschungsgemeinschaft (DFG) through
the DFG research group FOR5044 (Grant No. 426703838 \cite{FOR5044}; \url{https://www.for5044.de}). In addition, the TU Dresden team acknowledges the financial support by the Bundesministerium für Bildung und Forschung (BMBF, Federal Ministry of Education and Research, Germany, Project Grant Nos. 05K19ODA, 05K19ODB, and 05K22ODA), as well as by the DFG through projects CRC1415 (ID: 417590517), and the Würzburg-Dresden Cluster of Excellence ct.qmat (EXC 2147, ID: 390858490). Calculations for this research were conducted on the Lichtenberg high-performance computer of the TU Darmstadt and at the  H\"ochstleistungrechenzentrum Stuttgart (HLRS). The authors, furthermore, acknowledge the computational resources provided by the HPC Core Facility and the HRZ of the Justus-Liebig-Universit\"at Gie{\ss}en.

\section*{Data Availability}

The data that support the findings of this study are available from the corresponding author upon reasonable request.

\bibliography{literature_LNT}

\clearpage

\beginsupplement

\onecolumngrid

\section*{\label{sec:sup}Supplement}

\begin{figure*}
  \includegraphics[scale=0.2]{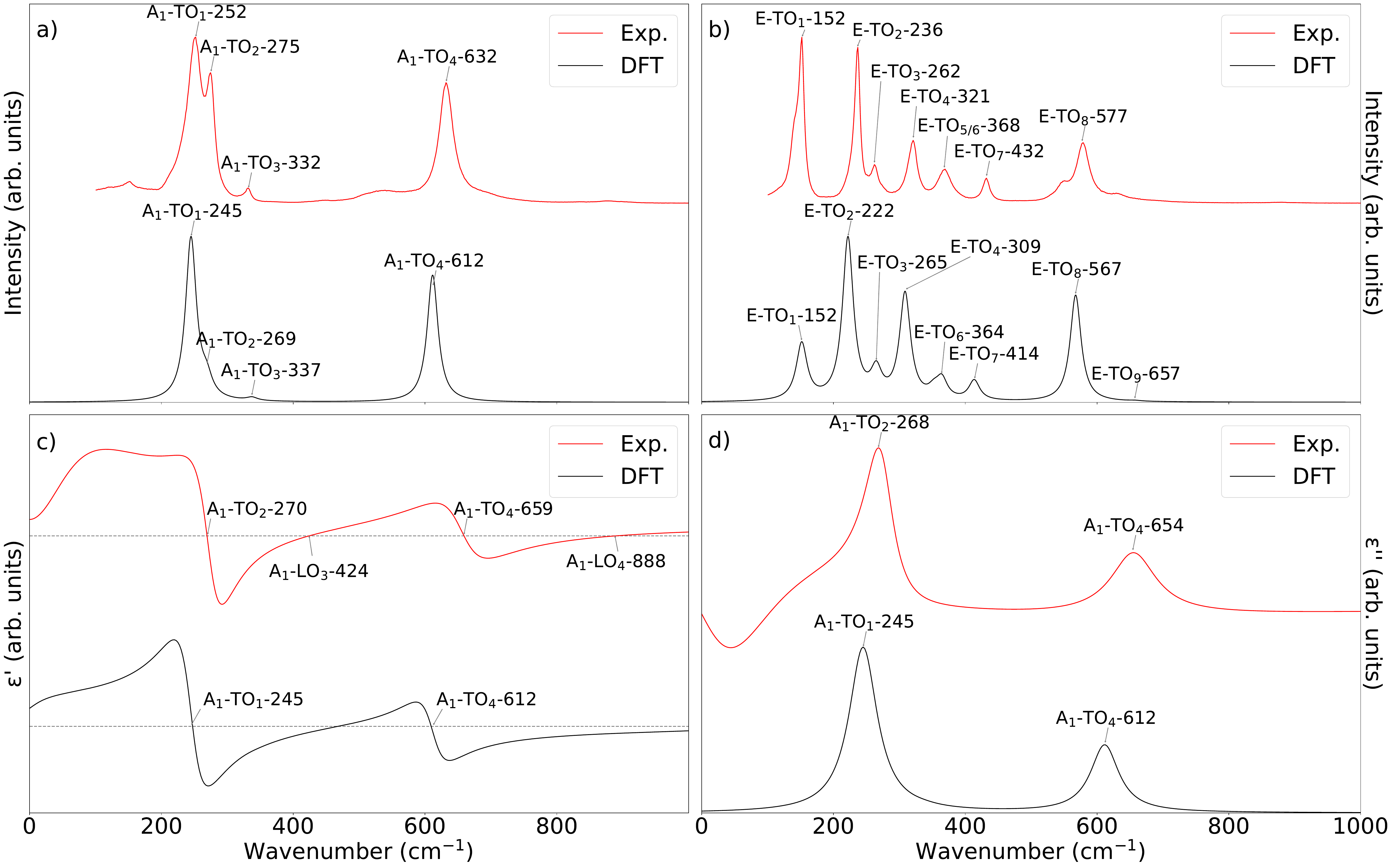}
  \caption{\label{fig:LNO_z_sup}Top: a) Raman spectra of LNO for polarization x(zz)$\overline{\rm{x}}$ (left) and b) x(zy)$\overline{\rm{x}}$ (right); Bottom: c) Real and d) imaginary dielectric functions for light polarized along the crystallographic z direction, extracted from FTIR spectra. All spectra are normalized with respect to their highest peak. In the Raman spectra, the A$_1$-TO and E-TO modes can be unambiguously assigned, by applying the symmetry considerations from Table \ref{tab:porto}. The A$_1$-LO modes can be extracted as zero crossings or poles in $\varepsilon'$. The redundancies in these four graphics serve as an appraisal for the measurements and calculations.}
\end{figure*}

\begin{figure*}
  \includegraphics[scale=0.2]{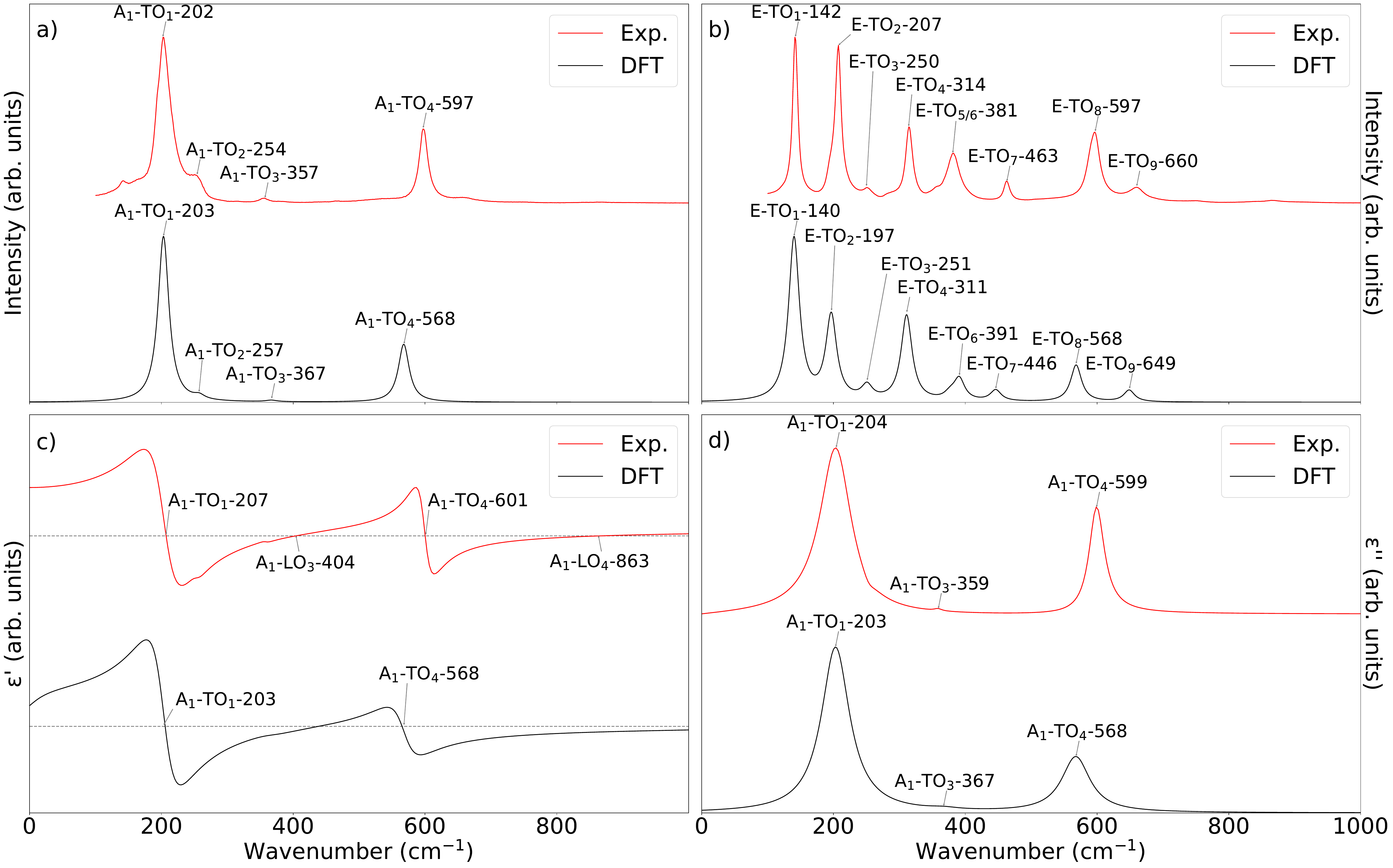}
  \caption{Top: a) Raman spectra of LTO for polarization x(zz)$\overline{\rm{x}}$ (left) and b) x(yz)$\overline{\rm{x}}$ (right); Bottom: c) Real and d) imaginary dielectric functions for light polarized along the crystallographic z direction, extracted from FTIR spectra. All spectra are normalized with respect to their highest peak. In the Raman spectra, the A$_1$-TO and E-TO modes can be unambiguously assigned, by applying the symmetry considerations from Table \ref{tab:porto}. The A$_1$-LO modes can be extracted as zero crossings or poles in $\varepsilon'$. The redundancies in these four graphics serve as an appraisal for the measurements and calculations.}
\end{figure*}

\begin{figure*}
  \includegraphics[scale=0.2]{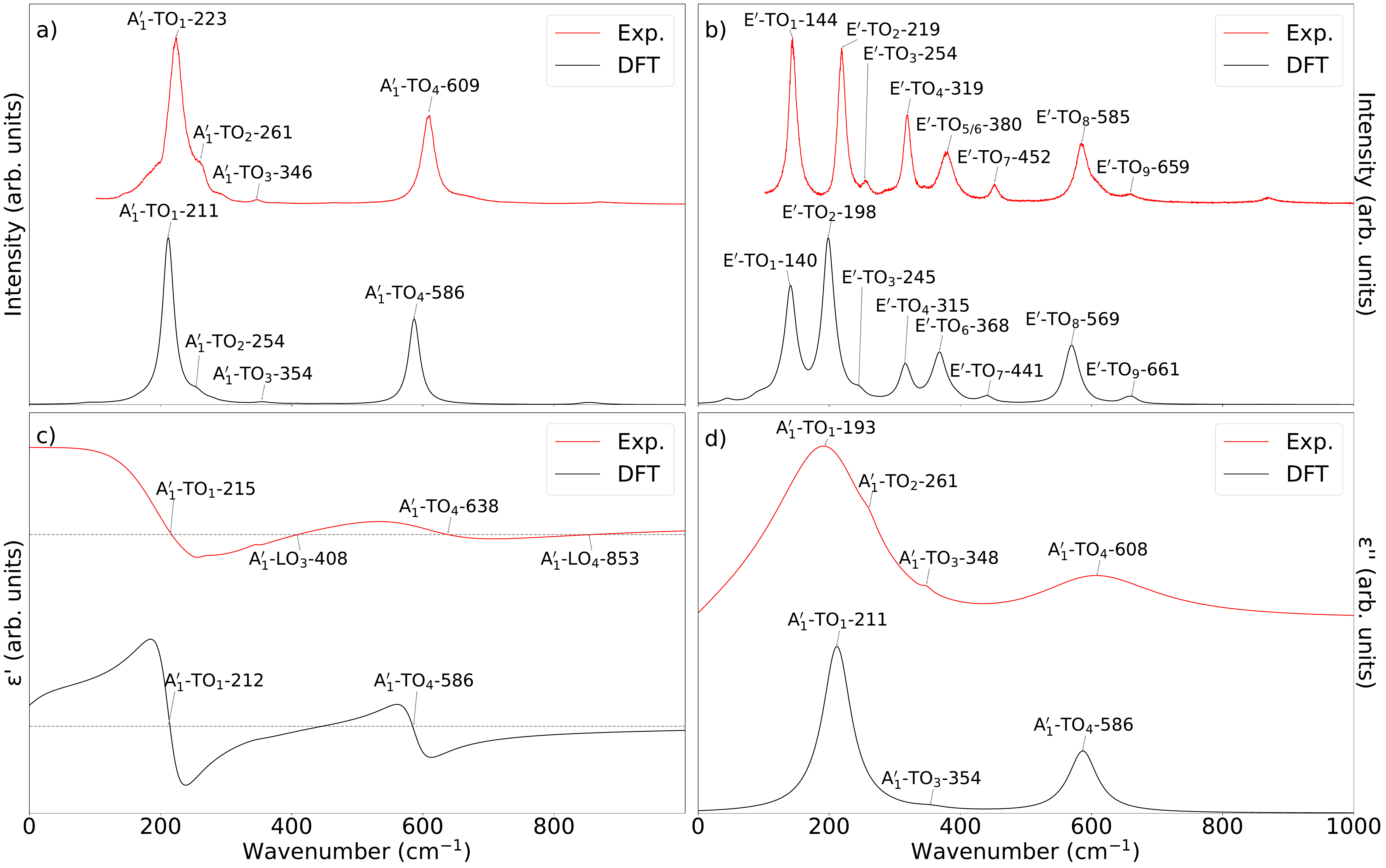}
  \caption{Top: a) Raman spectra of LNT for polarization x(zz)$\overline{\rm{x}}$ (left) and b) x(zy)$\overline{\rm{x}}$ (right); Bottom: c) Real and d) imaginary dielectric functions for light polarized along the crystallographic z direction, extracted from FTIR spectra. All spectra are normalized with respect to their highest peak. In the Raman spectra, the A$'_1$-TO and E$'$-TO modes can be unambiguously assigned, by applying the symmetry considerations from Table \ref{tab:porto}. The A$'_1$-LO modes can be extracted as zero crossings or poles in $\varepsilon'$. The redundancies in these four graphics serve as an appraisal for the measurements and calculations.}
\end{figure*}

\begin{figure*}
  \includegraphics[scale=0.2]{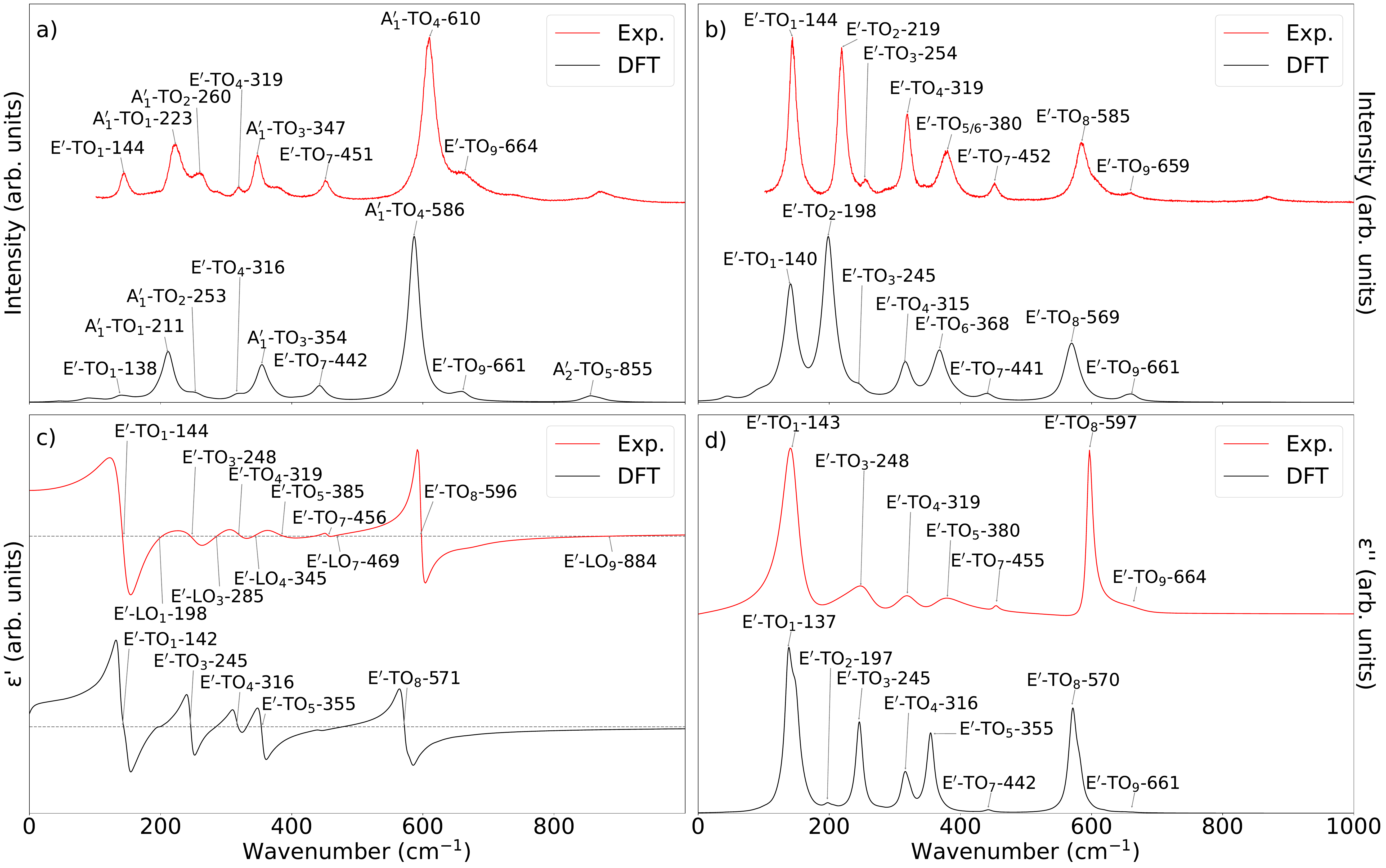}
  \caption{Top: a) Raman spectra of LNT for polarization x(yy)$\overline{\rm{x}}$ (left) and b) x(zy)$\overline{\rm{x}}$ (right); Bottom: c) Real and d) imaginary dielectric functions for light polarized along the crystallographic x direction, extracted from FTIR spectra. All spectra are normalized with respect to their highest peak. In the Raman spectra, the A$'_1$-TO and E$'$-TO modes can be unambiguously assigned, by applying the symmetry considerations from Table \ref{tab:porto}. The E$'$-LO modes can be extracted as zero crossings or poles in $\varepsilon'$. The redundancies in these four graphics serve as an appraisal for the measurements and calculations.}
\end{figure*}

\begin{table}
\caption{\label{tab:modes_LN} Calculated (DFPT) and measured mean value (Exp.) TO phonon modes of LiNbO$_3$ at the Brillouin zone center.
Available theoretical \cite{SimoRaman15} (DFT) data are reported for comparison. All frequencies in
cm$^{-1}$}
\begin{ruledtabular}
\begin{tabular}{llll}
Symmetry & DFPT & DFT & Exp. \\
\hline
A$_1$-TO$_1$ & 245 & 239 & 250 \\
A$_1$-TO$_2$ & 269 & 289 & 275 \\
A$_1$-TO$_3$ & 337 & 353 & 331     \\
A$_1$-TO$_4$ & 612 & 610 & 632 \\

A$_2$-TO$_1$ & 211 & &  \\
A$_2$-TO$_2$ & 293 & &  \\
A$_2$-TO$_3$ & 394 & &  \\
A$_2$-TO$_4$ & 438 & &  \\
A$_2$-TO$_5$ & 865 & &  \\

E-TO$_1$ & 152 & 148 & 151 \\
E-TO$_2$ & 222 & 216 & 236 \\
E-TO$_3$ & 265 & 262 & 265 \\
E-TO$_4$ & 309 & 323 & 321 \\
E-TO$_5$ & 352 & 380 & 360     \\
E-TO$_6$ & 364 & 391 & 367     \\
E-TO$_7$ & 414 & 423 & 431     \\
E-TO$_8$ & 567 & 579 & 579 \\
E-TO$_9$ & 657 & 667 & 660     \\
\end{tabular}
\end{ruledtabular}
\end{table}

\begin{table}
\caption{\label{tab:modes_LT} Calculated (DFPT) and measured mean value (Exp.) TO phonon modes of LiTaO$_3$ at the Brillouin zone center. Available theoretical \cite{SimoRaman15}  (DFT) data are reported for comparison. All frequencies in
cm$^{-1}$}
\begin{ruledtabular}
\begin{tabular}{llll}
Symmetry & DFPT & DFT & Exp. \\
\hline
A$_1$-TO$_1$ & 203 & 209 & 204 \\
A$_1$-TO$_2$ & 257 & 286 & 253 \\
A$_1$-TO$_3$ & 367 & 376 & 357 \\
A$_1$-TO$_4$ & 568 & 591 & 599 \\

A$_2$-TO$_1$ & 175 &    &         \\
A$_2$-TO$_2$ & 282 &    &         \\
A$_2$-TO$_3$ & 384 &    &         \\
A$_2$-TO$_4$ & 440 &    &         \\
A$_2$-TO$_5$ & 887 &    &         \\

E-TO$_1$   & 140 & 144 & 142 \\
E-TO$_2$   & 197 & 199 & 208     \\
E-TO$_3$   & 251 & 253 & 254 \\
E-TO$_4$   & 311 & 319 & 315 \\
E-TO$_5$   & 377 & 409 & 373     \\
E-TO$_6$   & 391 & 420 & 382 \\
E-TO$_7$   & 446 & 459 & 462 \\
E-TO$_8$   & 568 & 590 & 591 \\
E-TO$_9$   & 649 & 669 & 660 \\
\end{tabular}
\end{ruledtabular}
\end{table}

\begin{table}
\caption{\label{tab:LO_modes_LN}Calculated (DFPT) and measured mean value (Exp.) LO phonon modes of LiNbO$_3$ at the Brillouin zone center.
Available experimental \cite{Margueron2012} (Ref.) data are reported for comparison. All frequencies in cm$^{-1}$}
\begin{ruledtabular}
\begin{tabular}{llll}
Symmetry & DFPT & Exp. & Ref.  \\
\hline
A$_1$-LO$_1$ & 274 & 275 & 276 \\
A$_1$-LO$_2$ & 334 & 333 & 334 \\
A$_1$-LO$_3$ & 402 & 428 & 421  \\
A$_1$-LO$_4$ & 815 & 873 & 870 \\

E-LO$_1$ & 186 & 194 & 199 \\
E-LO$_2$ & 221 & 238 & 241 \\
E-LO$_3$ & 293 & 296 & 298 \\
E-LO$_4$ & 341 & 345 &     \\
E-LO$_5$ & 369 & 369 & 370 \\
E-LO$_6$ & 394 & 425 & 425 \\ 
E-LO$_7$ & 427 & 457 & 457 \\
E-LO$_8$ & 652 & 661 &     \\
E-LO$_9$ & 867 & 880 & 879 \\
\end{tabular}
\end{ruledtabular}
\end{table}

\begin{table}
\caption{\label{tab:LO_modes_LT}Calculated (DFPT) and measured mean value (Exp.) LO phonon modes of LiTaO$_3$ at the Brillouin zone center.
Available experimental \cite{Margueron2012} (Ref.) data are reported for comparison. All frequencies in cm$^{-1}$}
\begin{ruledtabular}
\begin{tabular}{llll}
Symmetry & DFPT & Exp. & Ref. \\
\hline
A$_1$-LO$_1$ & 258 & 254 & 259 \\
A$_1$-LO$_2$ & 356 & 354 & 356 \\
A$_1$-LO$_3$ & 393 & 405 & 408 \\
A$_1$-LO$_4$ & 820 & 865 & 863 \\

E-LO$_1$ & 177 & 194 & 195 \\
E-LO$_2$ & 197 & 206 & 211 \\
E-LO$_3$ & 277 & 279 & 282 \\
E-LO$_4$ & 343 & 344 &  \\
E-LO$_5$ & 391 & 381 & 383 \\
E-LO$_6$ & 438 & 453 & 454 \\
E-LO$_7$ & 461 & 472 & 476 \\
E-LO$_8$ & 648 & 662 & 660 \\
E-LO$_9$ & 820 & 865 & 864 \\
\end{tabular}
\end{ruledtabular}
\end{table}

\begin{table}
\caption{\label{tab:modes_LNT}Calculated (DFPT) and measured mean value (Exp.) phonon modes of LiTa$_{0.7}$Nb$_{0.3}$O$_3$ at the Brillouin zone center. The with $^*$ denoted frequencies are obtained by linearily interpolating the corresponding frequencies of the end compounds, as has been shown in Ref.~\cite{RuesPRB}. All frequencies in cm$^{-1}$}
\begin{ruledtabular}
\begin{tabular}{lllll}
Symmetry & \multicolumn{2}{c}{DFPT} & \multicolumn{2}{c}{Exp.} \\
& TO & LO & TO & LO \\
\hline
A$_1'$ & 211 & 263$^*$& 223 & 260\\
A$_1'$ & 260 & 348 & 261 & 346 \\
A$_1'$ & 354 & 401 & 346 & 414 \\
A$_1'$ & 586 & 841 & 609 & 869 \\

E$_1'$ & 140 & 180$^*$ & 144 & 193 \\
E$_2'$ & 198 & 202 & 220 & 215 \\
E$_3'$ & 245 & 284 & 256 & 287 \\
E$_4'$ & 316 & 337 & 319 & 344 \\
E$_5'$ & 371$^*$ & 384$^*$ & 377 & 366 \\
E$_6'$ & 385$^*$ & 439 & 379 & 445 \\
E$_7'$ & 442 & 456 & 452 & 467 \\
E$_8'$ & 569 & 649$^*$ & 586 & 664\\
E$_9'$ & 661 & 841 & 657 & 875 \\
\end{tabular}
\end{ruledtabular}
\end{table}

\end{document}